\documentclass[12pt]{article}

\usepackage{amssymb, amsmath}
\usepackage{graphicx}
\usepackage{cite}

\newcommand{\const}{\,{\rm const}\,}

\def\be{\begin{equation}}
\def\ee{\end{equation}}
\def\bea{\begin{eqnarray}}
\def\eea{\end{eqnarray}}

\title{ \bf{An infinite class of extremal horizons in higher dimensions}}

\author{Hari K. Kunduri$^a$\footnote{hkunduri@phys.ualberta.ca } \  and James Lucietti$^b$\footnote{j.lucietti@imperial.ac.uk } \\ \\
\small \sl $^a$ Theoretical Physics Institute,
Department of Physics, University of Alberta, \\ \small \sl Edmonton, AB, T6G 2J1, Canada
\\ \\ \small \sl $^b$ Theoretical Physics, Blackett Laboratory, Imperial College London, \\ \small \sl  London, SW7 2AZ, UK }

\date{}

\begin{document}

\maketitle

\begin{picture}(0,0)(0,0)
%\put(350, 320){}
\put(350, 310){Imperial/TP/2010/JL/01}
\put(350, 295){Alberta Thy 02-10}
\end{picture}

\vskip1.5cm

\begin{abstract}
We present a new class of near-horizon geometries which solve Einstein's vacuum equations, including a negative cosmological constant, in all even dimensions greater than four. Spatial sections of the horizon are inhomogeneous $S^2$-bundles over any compact K\"ahler-Einstein manifold. For a given base, the solutions are parameterised by one continuous parameter (the angular momentum) and an integer which determines the topology of the horizon. In six dimensions the horizon topology is either $S^2\times S^2$ or $\mathbb{CP}^2\# \overline{\mathbb{CP}^2}$. In higher dimensions the $S^2$-bundles are always non-trivial, and for a fixed base, give an infinite number of distinct horizon topologies.  Furthermore, depending on the choice of base we can get examples of near-horizon geometries with a single rotational symmetry (the minimal dimension for this is eight). All of our horizon geometries are consistent with all known topology and symmetry constraints for the horizons of asymptotically flat or globally  Anti de Sitter extremal black holes.
\end{abstract}

\newpage

\tableofcontents
\newpage

\section{Introduction}
 One of the classic results of four dimensional General Relativity is Hawking's horizon topology theorem~\cite{HawkingCMP, CW}. This states that spatial sections of the event horizon of an asymptotically flat black hole solution to Einstein's equations must be homeomorphic to $S^2$. This theorem is a key ingredient to the black hole uniqueness theorem, being the first logical step required to prove such a classification theorem for black holes.

For a variety of reasons, mainly stemming from String Theory and AdS/CFT (see e.g.~\cite{ERrev} for a clear account and references), the study of black hole solutions to higher dimensional General Relativity has recently attracted a great deal of attention. The classification of black hole solutions in higher than four dimensions is a difficult open problem. However,  it is possible to extend some of the ingredients which were used for $D=4$ to higher dimensions.  For example, Hawking's horizon topology theorem uses the two dimensional Gauss-Bonnet theorem in a crucial way and thus does not generalise straightforwardly. Nevertheless, Galloway and Schoen~\cite{GS} have established a generalisation which constrains the horizon topology of asymptotically flat black holes in the following manner: spatial sections of the event horizon, $\mathcal{H}$ (which are $D-2$ dimensional orientable and closed manifolds), must be positive Yamabe type\footnote{A compact manifold is positive Yamabe type if and only if it admits a positive scalar curvature metric.}. For $D=5$, so $\mathcal{H}$ is three dimensional, this constraint is strong enough to allow only $S^3$ (and quotients) and $S^1 \times S^2$ (and connected sums of these). In fact, explicit asymptotically flat black hole solutions are known for both of these topology types~\cite{MP, ER1, PS}. In $D \geq 6$, so $\textrm{dim}\, \mathcal{H} \geq 4$, the  complete list of positive Yamabe type manifolds is not known. However, it is clearly less of a constraint than in $D=5$.

There is in fact another way of constraining horizon topologies, as noted in~\cite{Reall}.  Suppose we have a black hole solution which is asymptotically flat or globally Anti de Sitter\footnote{By asymptotically globally AdS we mean the conformal boundary is $\mathbb{R}\times S^{D-2}$. We will not consider asymptotically locally AdS spacetimes, i.e. with conformal boundary $\mathbb{R}^{}\times X$ for more general $X$. For black holes with these asymptotics, $\mathcal{H}$ would have to be cobordant to $X$. Recall that another important case in the context of AdS/CFT is $X=\mathbb{R}^{D-2}$ in which case the known ``black hole" solutions have $\mathcal{H}=\mathbb{R}^{D-2}$.} (AdS). Now consider a spacelike hypersurface $\Sigma$ which intersects the future event horizon and conformal future infinity. The $D-1$ dimensional manifold $\Sigma$ has a boundary which is the disjoint union of $\mathcal{H}$ and $S^{D-2}$ (the sphere at infinity), i.e. it defines a cobordism between $\mathcal{H}$ and $S^{D-2}$. In fact since the manifolds in question all have an orientation induced from the spacetime orientation, then $\mathcal{H}$ and $S^{D-2}$ must be oriented cobordant. It is a standard result that two closed manifolds are (oriented) cobordant if and only if their corresponding Stiefel-Whitney and Pontryagin numbers are equal~\cite{MS, Wall}.  Since these numbers all vanish for spheres we deduce that $\mathcal{H}$ must have vanishing Stiefel-Whitney and Pontryagin numbers.

It is worth noting that topological censorship requires $\Sigma$ to be simply connected~\cite{CW, GSWW}. However, for $\textrm{dim} \; \mathcal{H} \geq 3$ this provides no extra constraint on the topology of $\mathcal{H}$ because given any oriented cobordism there must always exist a simply connected oriented cobordism \cite{Milnor1961}. In fact for $\textrm{dim } \mathcal{H}=3$ all the Stiefel-Whitney and Pontryagin numbers trivially vanish for any $\mathcal{H}$, and thus the existence of such cobordisms provides no constraint. However, for $\textrm{dim} \, \mathcal{H}  \geq 4$ the existence of a cobordism to a sphere does give non-trivial constraints on the topology, which is different to the positive Yamabe constraint. For example, $\mathbb{CP}^2$ is positive Yamabe but has non vanishing Stiefel-Whitney and Pontryagin numbers, whereas $T^4$ is zero Yamabe type but has vanishing Stiefel-Whitney and Pontryagin numbers. Therefore for $D>5$ the existence of such cobordisms provides a refinement of allowed horizon topologies for asymptotically flat and AdS black holes.

Now, as is well known, finding, and let alone classifying, black hole solutions is a difficult task. In this paper we are motivated by the question: what horizon topologies are actually realised by asymptotically flat and globally AdS black hole solutions in $D>5$?\footnote{Note that this question is still open for $D=5$ black holes.} As we have discussed above a necessary condition is they are oriented-cobordant to a sphere (or equivalently have vanishing Stiefel-Whitney and Pontryagin numbers), and, at least in the asymptotically flat case, positive Yamabe type. But are these conditions sufficient? This is a fundamental open problem towards the classification of higher dimensional black holes.

Interestingly, for extremal black holes, one can show that the full spacetime Einstein equations imply the metric induced on $\mathcal{H}$ satisfies an equation which depends only on intrinsic data on $\mathcal{H}$. Thus in a precise sense the Einstein equations on the horizon can be decoupled and solved separately. This is in fact intimately related to the existence of the so called near-horizon limit of the full black hole metric~\cite{Reall, CRT, KLR}. Therefore,  by studying the horizon equation one can learn about the possible horizon geometries and topologies for $\mathcal{H}$, without finding the full black hole metric. This is the approach we will take. It is worth emphasising that this method can allow one to rule out possible black horizon topologies, but not prove their existence (since given a near-horizon geometry there need not be a  corresponding black hole solution).

To get some insight into what one might expect, consider possible near-horizon geometries in Einstein-Maxwell theory. The reason for doing this is that it is easy to construct some simple static examples, i.e. the direct product AdS$_2 \times \mathcal{H}$, where the metric on $\mathcal{H}$ is positive Einstein, with a Maxwell field proportional to the volume form on AdS$_2$. For simplicity consider $D=6$, and thus in this example $\mathcal{H}$ is a positive Einstein closed 4-manifold. The classification problem for such spaces is a famous open problem in differential geometry. Only a few explicit examples of such Einstein spaces are known~\cite{Besse}: $S^4$ with the round metric, $\mathbb{CP}^2$ with the Fubini-Study metric, $S^2\times S^2$ with the standard product metric and $\mathbb{CP}^2 \# \overline{\mathbb{CP}^2}$ (i.e. $\mathbb{CP}^2$ with 1-point blown up) with the Page metric~\cite{Page} (which is cohomogeneity-1 and conformally K\"ahler). Existence of Einstein metrics has been established for $\mathbb{CP}^2 \# k \overline{\mathbb{CP}^2}$ with $2 \leq k \leq 8$ \footnote{The $k=2$ case is conformally K\"ahler whereas the rest are K\"ahler.}. This provides us with a host of Einstein metrics which can be used to give near-horizon geometries of exotic horizon topology. Out of this list only $S^4$, $S^2 \times S^2$ and $\mathbb{CP}^2 \# \overline{\mathbb{CP}^2}$ are actually cobordant\footnote{This can been seen from the fact that $\mathbb{CP}^2$ is the generator of the oriented cobordism group in four dimensions, which is in fact isomorphic to $\mathbb{Z}$.} to $S^4$. However, in view of the uniqueness theorem for static black holes with an electric field~\cite{GIS} only the $S^4$ case is expected to arise as a limit of a static asymptotically flat black hole (although note this theorem has only been proved for non-extremal black holes). Nevertheless, one might expect that rotating black holes (stationary and non-static case) could have $S^2\times S^2$ or $\mathbb{CP}^2 \# \overline{\mathbb{CP}^2}$ horizon topologies. This would be analogous to black rings in $D=5$ which must be rotating.

In this paper we construct an infinite family of vacuum (and also Einstein) near-horizon geometries in $D=2n+2 \geq 6$ which have non-spherical horizon topology. Spatial sections of the horizon $\mathcal{H}$ are smooth inhomogeneous $2n$-dimensional $S^2$-bundles over a compact K\"ahler-Einstein manifold $K$ (the Einstein metric on $K$ has positive curvature). Our ansatz is inspired by the local form of the horizon metric of certain even dimensional extremal Myers-Perry black holes (which can be written as fibrations over $\mathbb{CP}^{n-1}$), as well as certain Einstein metrics on complex line bundles over $K$~\cite{PP} (which include Page's metric~\cite{Page}).  For fixed $K$, our solutions are parameterised by a continuous parameter $L>0$ and an integer $m>p$ which specifies the topology of $\mathcal{H}$ (where $p$ is an integer associated to the K\"ahler-Einstein base called the Fano index).  The isometry group of our near-horizon geometries is $SO(2,1)\times U(1)\times G$ where $G$ is the isometry group of $K$.  Note that by construction, the local form of our solutions (with $K=\mathbb{CP}^{n-1}$) also contain the near-horizon geometry of the extremal Myers-Perry-(AdS) black holes (which have $\mathcal{H}=S^{2n}$) with all angular momenta equal as a special case.

In $D=6$ (so $n=2$), these solutions give smooth cohomogeneity-1 horizon geometries for $\mathcal{H}=S^2 \times S^2$ (if $m$ is even) and $\mathcal{H}=\mathbb{CP}^2\# \overline{\mathbb{CP}^2}$ (if $m$ is odd). Note that these horizon metrics are \emph{not} the Einstein ones for $S^2\times S^2$ or $\mathbb{CP}^2 \# \overline{\mathbb{CP}^2}$ discussed above in the context of static near-horizon geometries in Einstein-Maxwell -- in particular our $S^2\times S^2$ metric is not even a product metric. As discussed above both of these manifolds are cobordant to $S^4$ and are positive Yamabe and therefore candidates as horizons of black holes. Therefore we will discuss the possibility that there are asymptotically flat or globally AdS vacuum black hole solutions with such horizon topologies, and that the solution we have is the near-horizon limit of an extremal black hole of this kind.

In higher dimensions (i.e. $n>2$),  the new near-horizon geometries we find are all non-trivial $\mathcal{H}$-bundles over AdS$_2$, with $\mathcal{H}$ itself always a non-trivial $S^2$-bundle over any compact K\"ahler-Einstein space. Strikingly, for a fixed K\"ahler-Einstein base, the topology of the horizon (i.e. the $S^2$-bundle over $K$) is different for each value of the integer $m$. Therefore, we have an {\it infinite} discrete class of horizon topologies (in contrast to the $n=2$ case above). Furthermore, as we explain later, any $S^2$-bundle over a compact manifold is guaranteed to be cobordant to a sphere, and any $S^2$-bundle over a compact base with positive Ricci curvature must be of positive Yamabe type. Therefore, all our horizon topologies are cobordant to $S^{2n}$ and positive Yamabe type, and are thus all consistent with the topological restrictions discussed above for the horizons of asymptotically flat and globally AdS black holes.

If the K\"ahler-Einstein base space $K$ is toric (i.e. admits $U(1)^{n-1}$ isometry) then these near-horizon geometries have $SO(2,1)\times U(1)^n$ isometry where $n=[(D-1)/2]$ is the rank of $SO(D-1)$. Interestingly, if one chooses the K\"ahler-Einstein space to have no isometries\footnote{The minimal dimension for such near-horizon geometries is $8$ (i.e. $n=3$ so dim $K=4$). Explicit examples for dim $K=4$ are given by the del Pezzo surfaces $dP_k$ for $4\leq k \leq 8$, i.e. $\mathbb{CP}^2$ with $k$ points blown up in general positions. }, we get examples of near-horizon geometries with isometry exactly $SO(2,1)\times U(1)$, i.e. just one rotational isometry $U(1)$ and no more. This is interesting as it has been conjectured that in view of the higher dimensional version of the rigidity theorem~\cite{HIW,IM}, there should be stationary black hole solutions with $\mathbb{R} \times U(1)$ symmetry~\cite{Reall}. Therefore we will discuss the interesting possibility that our near-horizon geometries are near-horizon limits of extremal black holes with this minimal amount of rotational symmetry.

It is worth emphasising that our solutions possess no more abelian rotational symmetry than is allowed for asymptotically flat or globally AdS spacetimes: $U(1)^{[(D-1)]/2}$ (and this is saturated when $K$ is toric). Indeed this was the motivation for focusing on the class of near-horizon geometries considered in this paper. However, they do not constitute the most general possibility with $U(1)^{[(D-1)/2]}$ rotational symmetry. Indeed the classification of near-horizon geometries with $U(1)^{[(D-1)]/2}$ symmetry is an interesting open problem out of reach with current methods.  However, when the K\"ahler-Einstein space is chosen to be homogeneous our near-horizon geometries are cohomogeneity-1.  It is then plausible (at least for $K=\mathbb{CP}^{n-1}$) that our solutions are the most general cohomogeneity-1 near-horizon geometries with a maximal abelian isometry group $U(1)^{[(D-1)/2]}$, although we have not proved this.

The analogous classification problem in $D=4$~\cite{Haj,LP,KL1} and $D=5$~\cite{KL1} has been solved (including a Maxwell field and a cosmological constant in $D=4$~\cite{LP, KL1,KL2}) -- i.e. the classification of near-horizon geometries with $U(1)$ and $U(1)^2$ rotational symmetry respectively. Crucially, these near-horizon classifications have been used recently to prove uniqueness theorems for extremal Kerr~\cite{AHMR, FL, CN} and Kerr-Newman~\cite{AHMR, CN} as well as a $D=5$ generalisation~\cite{FL}. For $D>5$ there is another possible generalisation of these 4d and 5d problems. That is, the classification of near-horizon geometries with $U(1)^{D-3}$ rotational symmetry. This has also been solved~\cite{HI}, however it is worth emphasising that for $D>5$ such near-horizon geometries cannot be near-horizon limits of asymptotically flat black holes since they have too many commuting rotational isometries -- instead they would arise as near-horizon limits of Kaluza-Klein (KK) black holes which are uniform in the KK direction.

The organisation of this paper is as follows. In Section 2 we present the simplest example of our solutions, a Ricci-flat near-horizon geometry in six dimensions with horizon topology either $S^2 \times S^2$ or $\mathbb{CP}^{2}\#\overline{\mathbb{CP}}^{2}$. This short summary is intended for readers who wish to avoid our analysis in detail. Section 3 presents the derivation of our new near-horizon solutions. The global analysis of the resulting horizon geometries is given in Section 4. Section 5 presents a derivation of the physical properties of these solutions. Finally, in Section 6 we gather the preceding results and consider the possibility that these near-horizon geometries extend to extremal, asymptotically flat or asymptotically globally AdS black hole solutions. We conclude with a Discussion. Some useful technical results are collected in the Appendices.

\section{Summary of six dimensional Ricci flat near-horizon geometry}
\label{summary}
In this section we will summarise the six dimensional Ricci flat  near-horizon geometry we have found. We have also constructed analogous solutions with a negative cosmological constant and also in any even dimension. However, the Ricci flat six dimensional case is the simplest and thus a good example to illustrate our more general class of solutions. We will not give any derivations in this section, however we will make the presentation self-contained. In the subsequent sections we will provide complete derivations of the general class in all even dimensions including a cosmological constant.

The $D=6$ vacuum solution takes the explicit form
\bea
&&ds^2= (\xi_m+x^2)  \left[  -\frac{4r^2}{L^2} dv^2 +2dvdr  +\frac{L^2 (1-x^2) dx^2}{(4-m^2x^2)\left(\xi_m - \frac{4 x^2}{3m^2}\right)}  \right]   \\ \nonumber
&& \qquad + L^2 \left[  \frac{(4-m^2x^2)\left(\xi_m - \frac{4 x^2}{3m^2}\right)}{(\xi_m+x^2)(1-x^2)} \left(d\phi+ \frac{1}{2}\cos\theta d\chi  +2 \sqrt{\xi_m} rdv \right)^2  +\frac{1}{4}(1-x^2)( d\theta^2+\sin^2\theta d\chi^2) \right]
\eea
where
\be
\xi_m =\frac{4}{3} \left( \frac{3-\frac{4}{m^2}}{4+m^2}\right)
\ee
where $L>0$ and $m>2$ is an integer. The coordinate ranges are $-2/m \leq x \leq 2/m$, $0\leq \theta \leq \pi$, $\phi \sim \phi +2\pi/m$, $\chi\sim\chi+2\pi$  and $(v,r)$ can take any value (the horizon is at $r=0$).  Cross sections of the horizon $\mathcal{H}$, are homeomorphic to $S^2 \times S^2$ if $m$ is even or $\mathbb{CP}^2 \# \overline{\mathbb{CP}}^2$ if $m$ is odd. Note that the only other way the local form of the horizon metric can be extended to a smooth metric on a compact manifold is if $m=2$ and $\phi\sim \phi+2\pi$, which gives $\mathcal{H}=S^{4}$ and corresponds to the near-horizon limit of extremal 6d Myers-Perry with all angular momenta equal.

The area and Komar angular momentum, defined with respect to the rotational Killing field $m^{-1}\partial_\phi$ (since this has orbits with canonical period $2\pi$), are:
\bea
&&A(\mathcal{H})= \frac{8\pi^2 L^{4}}{3 m^2} \left( 3 -\frac{4}{m^2} \right)  \\
&& J= \pm \frac{\pi L^4}{2G} \frac{\sqrt{\xi_m}}{m} \left(1+\frac{4}{m^2} \right)
\eea
and therefore
\be
A(\mathcal{H})= 4\pi G \, m\sqrt{\xi_m} |J|  \; .
\ee
It is interesting to compare, for fixed $J$, the area of the new horizons $\mathcal{H}$ to the spherical topology case (\ref{MParea}). This can be expressed as
\be\label{ratio6d}
\frac{A(\mathcal{H})}{A(S^4)} = m\sqrt{3 \xi_m} = 2m\sqrt{ \frac{3-\frac{4}{m^2}}{4+m^2}}
\ee and therefore
\begin{equation}
2 < \frac{A(\mathcal{H})}{A(S^4)} < 2\sqrt{3}
\end{equation}
where the inequalities follow from the fact that~(\ref{ratio6d}) is a monototically increasing function of $m$ and $m>2$.

In section \ref{BH} we discuss the possibility that such near-horizon geometries arise as near-horizon limits of yet to be found asymptotically flat extremal black hole solutions.

\section{Construction of near-horizon geometries}
\subsection{Near-horizon equations}
\label{sec:NHequations}
We will assume that the event horizon of a stationary extremal black hole solution must be a Killing horizon of a Killing vector field $V$. In a neighbourhood of such a Killing horizon we can always introduce Gaussian  null coordinates~\cite{IM} $(v,r,x^A)$ such that $V = \partial / \partial v$, the horizon is at $r=0$ and $x^A$ are coordinates on a spatial section of the horizon $\mathcal{H}$ (which of course is $D-2$ dimensional). We will assume that $\mathcal{H}$ is an oriented  compact manifold without boundary. Near the extremal Killing horizon, the space-time metric in these coordinates reads
\be
 ds^2 = r^2 F(r,x) dv^2 + 2dvdr + 2rh_A(r,x) dvdx^A + \gamma_{AB}(r,x)dx^A dx^B  \; .
\ee
The near-horizon limit \cite{Reall,KLR} is obtained by taking the limit $v \to v/\epsilon, \ r \to \epsilon r$ and $\epsilon \to 0$. The resulting metric is
\be
\label{gNH}
 ds^2 = r^2F(x)dv^2 + 2dvdr + 2r h_A(x)dv dx^a + \gamma_{AB}(x)dx^A dx^B \; ,
\ee
where $F, h_A,  \gamma_{AB}$ are a function, a one-form, and a Riemannian metric respectively, defined on $\mathcal{H}$.

In this paper we will be interested in finding near-horizon geometry solutions to Einstein's vacuum equations $R_{\mu\nu}=\Lambda g_{\mu\nu}$. We will be mainly focused on $\Lambda \leq 0$.  One can prove (see e.g. ~\cite{KL1}) that these spacetime equations for a near-horizon geometry are in fact equivalent to the following set of equations on $\mathcal{H}$:
\be
\label{Riceq}
 R_{AB} = \frac{1}{2}h_A h_B - \nabla_{(A} h_{B)}    +\Lambda\gamma_{AB}
\ee
with the function $F$ determined by
\be
\label{Feq}
F = \frac{1}{2}h_Ah^A - \frac{1}{2}\nabla_A h^A +\Lambda \; ,
\ee
where $R_{AB}$ and $\nabla$ are the Ricci tensor and the covariant derivative of the metric $\gamma_{AB}$. In particular, (\ref{Riceq}) is the $AB$ component of the Einstein equations, (\ref{Feq}) is the $vr$ component, all written covariantly on $\mathcal{H}$. It can be shown that the rest of the Einstein equations are satisfied as a consequence of the above set of equations.

Before moving on we note that static near-horizon geometries of this kind (which are equivalent to $dh=0$) have been classified in~\cite{CRT}. It was found that for $\Lambda \leq 0$ the only solution is $F=\Lambda$, $h_A=0$ and $R_{AB}=\Lambda \gamma_{AB}$. In this paper we will focus on the non-static case.

\subsection{A class of near-horizon geometries in even dimensions}
We will consider $(2n+2)$-dimensional near-horizon geometries (so dim $\mathcal{H}=2n$)  of the form
\bea
 \label{gamma}
\gamma_{AB}dx^Adx^B &=&L^2\left[  A(\rho)^2 \bar{g}_{ab} d\bar{x}^a d\bar{x}^b + d\rho^2+ B(\rho)^2 (d\phi +\sigma)^2 \right] \\
h_Adx^A &=&  \Gamma(\rho)^{-1} [ k(\rho) B(\rho)^2 (d\phi+\sigma) - \Gamma'(\rho)d\rho ]  \label{h}
\eea
where $\bar{g}_{ab}$ is a $(2n-2)$-dimensional K\"ahler-Einstein metric on a base manifold $K$, normalised to $\bar{R}_{ab}=2n \bar{g}_{ab}$, with K\"ahler form $J=\frac{1}{2}d\sigma$ and $\bar{x}^a$ some set of coordinates on $K$. As we will see later, compactness of $\mathcal{H}$ requires the coordinates $\rho \in  [\rho_1, \rho_2 ]$,  $\phi\sim \phi+\Delta \phi$ and that $K$ be compact. The constant $L$ is taken to have dimensions of length and is introduced for later convenience.  The isometry group of $\gamma_{AB}$ is $G\times U(1)$ where $G$ is the isometry group of $\bar{g}_{ab}$ and the $U(1)$ is generated  by the Killing field $\partial /\partial \phi$. The 1-form $h$ is chosen to respect the $G\times U(1)$ symmetry too, so the total near-horizon geometry also has this symmetry.  In this parameterisation there is a scaling freedom,
\be
\label{scale}
(\rho, A, B, k, L) \to (s\rho, sA, sB, s^{-2}k, s^{-1}L)
\ee
where $s \neq 0$ is a constant, which leaves the near-horizon data invariant.

Let us explain our motivation for studying this class of near-horizon geometries.  One reason comes from choosing the base space $K$ to be a homogeneous space. Then our ansatz is in fact the most general horizon geometry with $G\times U(1)$ isometry group whose principle orbits are $U(1)$ bundles over a homogeneous base $K$ (even if $K$ is not K\"ahler-Einstein). The $(\rho, \phi)$  are coordinates valid on the principal orbits. Furthermore in this case $h$ is also the most general 1-form invariant under $G\times U(1)$ provided there is a unique  -- up to homothety -- homogeneous Einstein metric on $K$~\cite{FKLR} (in the K\"ahler case there is in fact a unique homogeneous K\"ahler-Einstein metric~\cite{Besse}). Note that if $K$ is homogeneous both the horizon and near-horizon geometries are cohomogeneity-1.

The case of most interest to us when $K$ is homogeneous is $K=\mathbb{CP}^{n-1}$, which has a unique K\"ahler-Einstein metric $\bar{g}_{ab}$ given by the Fubini-Study metric. This is indeed a homogeneous metric on $\mathbb{CP}^{n-1}$ with $G=SU(n)$, in which case the isometry group of the horizon geometry is $SU(n)\times U(1)$ with cohomogeneity-1 orbits.  This class of near-horizon geometries includes those of the extremal Myers-Perry black holes with all angular momenta equal~\cite{FKLR}. If instead we choose other K\"ahler-Einstein spaces $K$ with less symmetry  these metrics are no longer cohomogeneity-1  -- we will also be interested in this possibility (although then the above is not necessarily the most general near-horizon geometry with such symmetries). Another important point is that if $K$ is toric (i.e. has $U(1)^{n-1}$ symmetry) then the maximal abelian subgroup of $G\times U(1)$ is $U(1)^{[(D-1)]/2}$ which is the maximal abelian subgroup of the rotation group $SO(D-1)$ for asymptotically flat or globally AdS spacetimes.

We will now turn to solving equation (\ref{Riceq}). It is convenient to introduce a set of vielbeins\footnote{To avoid a proliferation of indices we will use the same symbols for coordinate and vielbein indices.} $e^A$  for the metric $\gamma_{AB}$:
\be
e^0=Ld\rho, \qquad e^a= LA \bar{e}^a, \qquad e^{2n-1}=LB (d\phi +\sigma)
\ee
where $a=1,\cdots 2n-2$ and $\bar{e}^a$ are vielbeins for $\bar{g}= \bar{e}^a \bar{e}^a$.
The Ricci tensor of (\ref{gamma}) in this basis is diagonal with
\bea\label{Ricci}
&&R_{00}= \frac{1}{L^2} \left[-\frac{2(n-1)A''}{A} - \frac{B''}{B} \right], \\  &&R_{2n-1 \, 2n-1} =\frac{1}{L^2}\left[ -\frac{B''}{B}+2(n-1) \left( \frac{B^2}{A^4} - \frac{A'B'}{AB} \right) \right] \\
&&R_{ab}=\left[ \frac{2n}{A^2} -\frac{A''}{A}- \frac{(2n-3) A'^2}{A^2} -\frac{2B^2}{A^4} - \frac{A'B'}{AB} \right] \frac{\delta_{ab}}{L^2} \; .
\eea Details of the calculation are given in Appendix \ref{Curvature}.
The source term $S_{AB}\equiv \frac{1}{2}h_A h_B- \nabla_{(A} h_{B)} +\Lambda \gamma_{AB}$ in the vielbein basis reads
\begin{eqnarray}
&&S_{00} = \frac{1}{L^2} \left( \frac{\Gamma''}{\Gamma}-\frac{\Gamma'^2}{2\Gamma^2} \right)+\Lambda, \qquad S_{2n-1 \, 2n-1}=  \frac{1}{L^2} \left( \frac{k^2 B^2}{2\Gamma^2}+ \frac{B' \Gamma'}{B \Gamma} \right) +\Lambda, \nonumber \\  &&S_{0 \, 2n-1} = - \frac{B k'}{L^2 \Gamma} \qquad S_{ab}=\frac{\Gamma' A'}{ \Gamma A} \frac{\delta_{ab}}{L^2} +\Lambda \delta_{ab} \; .
\end{eqnarray}
It immediately follows that the $0\, 2n-1$ component of equation (\ref{Riceq}) implies $k(\rho)=\kappa$ a constant.  We will assume $\kappa \neq 0$, otherwise the near-horizon geometry is static.

For solving the rest of the equations it is convenient to define a new coordinate $x$ by $x'(\rho)= B({\rho})$. Note that the coordinate is valid wherever $B \neq 0$ (which occurs on the principal orbits), and is defined up to the scaling freedom  $x \to s^2 x$ (inherited from (\ref{scale})), as well as $x\to -x$ and $x \to x+\const$. We will use these to simplify the solution.  For later reference it is worth noting that $(dx)^2=B^2/L^2$. From now on we will be treating everything as a function of $x$ and will denote $d/dx =\; '$.

Firstly, if one subtracts the $2n-1\, 2n-1$ component of equation (\ref{Riceq}) from the $0 \, 0$ component of equation (\ref{Riceq}) one gets
\be
\label{Aeq}
-\frac{2(n-1)}{A} \left( A''+ \frac{1}{A^3} \right) = \frac{1}{2\Gamma} \left( 2\Gamma'' - \frac{\Gamma'^2}{\Gamma} - \frac{\kappa^2}{\Gamma} \right) \; .
\ee
We will determine the most general solution for which the left and right sides of the above equation vanish {\it separately}.  It is not obvious that there must be solutions of this form, but as we shall see there are; it is also not necessarily the case that all solutions must be of this form.\footnote{In fact, in the $\Lambda=0$ case, it is easy to show there are solutions not in this class, e.g. there is a solution $B^2=nA^2$ and $\Gamma=|\kappa |  A^2/ (2 \sqrt{n-1})$ where $A$ is constant. If $K=\mathbb{CP}^{n-1}$ this is the near-horizon geometry of the direct product of an odd dimensional extremal Myers-Perry black hole with all angular momenta equal, and a line.} It is a guess inspired by the form of 4d near-horizon geometries~\cite{KL1,KL2} and even dimensional inhomogeneous Einstein spaces~\cite{PP}.  Therefore we should emphasise that we have not necessarily classified all solutions of the form (\ref{gamma}) and (\ref{h}).

Setting the RHS of (\ref{Aeq}) to zero gives an equation for $\Gamma$ which is identical to that which occurs for 4d near-horizon geometries~\cite{KL1,KL2}. Its general solution is
\be
\Gamma(x)= \frac{\kappa^2}{\beta}+ \frac{\beta (x-x_0)^2}{4}
\ee
where $\beta>0$ and $x_0$ are integration constants. For later convenience we will introduce a positive constant $\xi>0$ defined by
\be
\label{xidef}
\kappa^2=\frac{\beta^2 \xi}{4}
\ee
in terms of which
\be
\label{Gammax0}
\Gamma(x)= \frac{\beta}{4} [ \xi+ (x-x_0)^2 ] \; .
\ee
Setting the LHS of (\ref{Aeq}) to zero gives
\be
\label{Aeq2}
A''+A^{-3}=0 \;,
\ee
which in fact has two different families of solutions. Notice that (\ref{Aeq2}) implies $A$ is a non-constant function, a fact we will use below. We have thus determined the functions $\Gamma(x),A(x)$. Before solving for $A$ explicitly, we will now use (\ref{Aeq2}) to simplify the other field equations.

It is clear now that the $ab$ component of (\ref{Riceq}) is a first order equation for the remaining function $B$. In fact it is convenient to introduce the function
\be
\label{Ptildedef}
\tilde{P}(x)\equiv  B^2 \Gamma A^{2(n-1)} \; .
\ee
Then the $ab$ component of (\ref{Riceq}) can be simplified, using (\ref{Aeq2}), resulting in
\be
\label{Ptildeeq}
\frac{d}{dx} \left( \frac{\tilde{P}}{A {A'}} \right)= \frac{\Gamma A^{2(n-1)}( 2n-\lambda A^2)}{A^2 {A'}^2}
\ee
where $\lambda \equiv \Lambda L^2$ is dimensionless.

It is now sufficient to impose one other component of the field equations, say the $00$ component. We find that the $00$ component of (\ref{Riceq}) is (without using (\ref{Aeq2}))
\be
\tilde{P}'' - \left( \frac{\Gamma'}{\Gamma} +\frac{2(n-1) A'}{A} \right) \tilde{P}' + \left( \frac{2(n-1)A''}{A} + \frac{2(n-1) {A'}^2}{A^2} + \frac{\Gamma''}{\Gamma} \right) \tilde{P} =-2\lambda\Gamma A^{2(n-1)} \; .
\ee
Using (\ref{Aeq2}) this can be rewritten as
\bea
&&\frac{1}{AA'}\frac{d}{dx} \left[ A^2 {A'}^2 \frac{d}{dx}\left( \frac{\tilde{P}}{A {A'}} \right)\right] - \left( \frac{\Gamma'}{\Gamma} +\frac{2(n-1) A'}{A} \right) A{A'} \frac{d}{dx}\left( \frac{\tilde{P}}{A {A'}} \right)  \nonumber \\
&& \qquad + \left[ \frac{A A' \Gamma''}{\Gamma} + \left( \frac{1}{A^2} - {A'}^2 \right)\frac{\Gamma'}{\Gamma} \right] \frac{\tilde{P}}{AA'}=-2\lambda \Gamma A^{2(n-1)}  \; .  \label{00eq}
\eea
Now, substituting (\ref{Ptildeeq}) into (\ref{00eq}) results in many cancellations, and simplifies to
\be
\label{consistency}
\Gamma'' + \frac{1}{AA'} \left( \frac{1}{A^2}- {A'}^2 \right) \Gamma'=0  \; .
\ee
To summarise, we have shown that if we impose the ansatz (\ref{Aeq2}), then the field equations are equivalent to solving (\ref{Ptildeeq}), (\ref{consistency}) with (\ref{Gammax0}). Thus, this system of equations is overdetermined and it is not obvious there exist solutions. In fact as shall now see,  one class of solutions to (\ref{Aeq2}) leads to a full solution for this system.

First note that (\ref{Aeq2}) implies ${A'}^2-A^{-2}=\const$. In fact this constant must be nonzero, otherwise equation (\ref{consistency}) implies $\Gamma''=0$ which is inconsistent with (\ref{Gammax0}). Therefore we can write ${A'}^2-A^{-2}=-\epsilon \alpha^{-1}$ for some constant $\alpha > 0$ and $\epsilon=\pm 1$. Integrating\footnote{Note that the other solution $A^2=\epsilon \alpha$ to the first order equation ${A'}^2-A^{-2}=-\epsilon \alpha^{-1}$ does not solve the original second order equation ${A''}+A^{-3}=0$ and thus must be discarded.} one gets $A^2=\epsilon (\alpha -\alpha^{-1} x^2)$ where we have used the translation freedom in the definition of $x$ to fix the integration constant. Furthermore, using the scaling freedom (\ref{scale}) (with $s=\sqrt{\alpha}$) the solution can be written as
\be
\label{Asq}
A^2=\epsilon (1-x^2)  \;  .
\ee
Now, the equation (\ref{consistency}), using (\ref{Gammax0}) and (\ref{Asq}),  is satisfied if and only if $x_0=0$. Therefore $\Gamma$ is simply
\be
\label{Gamma}
\Gamma= \frac{\beta}{4} ( \xi +x^2)  \;.
\ee
Finally, substituting (\ref{Asq}) and (\ref{Gamma}) into (\ref{Ptildeeq})  gives
\be
\label{Peq}
\frac{d}{dx} \left( \frac{P(x)}{x} \right) = \frac{(\xi+x^2) (1-x^2)^{n-1}[ \lambda \epsilon (1-x^2)-2n]}{x^2}
\ee
where for convenience we have defined
\be
\label{Pdef}
P(x) \equiv 4\beta^{-1} \epsilon^n \tilde{P}(x) \; .
\ee
To summarise, we have  found a solution to the near-horizon equations (\ref{Riceq}) of the form (\ref{gamma}) and (\ref{h}), with $A^2$ and $\Gamma$ given by (\ref{Asq}) and (\ref{Gamma}) respectively, and $B^2$ determined up to a first order ODE for $P$ (\ref{Peq}) where $P$ is defined by (\ref{Pdef}) and (\ref{Ptildedef}).

To integrate the ODE for $P(x)$ (\ref{Peq}) explicitly it is convenient to define the polynomial
\be
\label{Qpol}
Q_n(u,\xi) \equiv  \sum_{l=0}^{n} {}^nC_l \left[ \frac{l-(n-l)\xi  }{n(2l-1)} \right] (-1)^l u^{l}
\ee
which satisfies
\be
\frac{d}{dx} \left( \frac{Q_n(x^2,\xi)}{x} \right) = - \frac{(\xi+x^2)(1-x^2)^{n-1}}{x^2} \; .
\ee
It is worth noting for later reference that
\be
\label{QGegen}
Q_n(x^2,\xi)= \frac{(n-1)! \sqrt{\pi}}{2 \Gamma(n+\frac{1}{2})} \left[ x C_{2n-1}^{(-n+\frac{1}{2})}(x)+ \xi C_{2n-2}^{(-n+\frac{1}{2})}(x) \right]
\ee
where $C^{(\alpha)}_n(x)$ are the Gegenbauer polynomials.
It then follows that the general solution to (\ref{Peq}) is:
\bea
&&P(x)= 2n Q_n(x^2,\xi) -\epsilon\lambda Q_{n+1}\left(x^2,\xi \right)+cx \label{P}
\eea
where $c$ is an integration constant. This explicit expression will be useful for later analysis.

We have thus determined the horizon data $\gamma_{AB},h_A$ in the coordinate system $(x,\phi, \bar{x}^a)$.
We may now evaluate the remaining near-horizon data, namely the function $F$ given by (\ref{Feq}). Using the explicit form of $A(x)$ and $\Gamma(x)$, together with (\ref{Peq}) (i.e. we do not need the explicit expression for $P(x)$) we find it can be written as:
\be\label{F}
F= \frac{A_0}{\Gamma} + \frac{\kappa^2 B^2}{L^2 \Gamma^2}
\ee
for a constant $A_0$ given by
\be
\label{A0}
A_0=-\frac{\epsilon n \beta}{2L^2}+\Lambda \left( \frac{\kappa^2}{\beta}+ \frac{\beta}{4} \right)= \frac{\beta}{4L^2} \left[ -2n\epsilon +\lambda(1+\xi) \right]
\ee
where in the second equality we have used (\ref{xidef}).
The significance of this particular form for $F$ is revealed by changing the radial variable in the full near-horizon geometry to $r \to \Gamma r$ which allows one to write it as
\be\label{NHGamma}
ds^2=\Gamma(x) [A_0 r^2 dv^2+2dvdr ]+ L^2 \left[  A(x)^2 \bar{g}_{ab} d\bar{x}^a d\bar{x}^b+ \frac{dx^2}{B(x)^2} + B(x)^2 \left(d\phi+\sigma+ \frac{\kappa  rdv}{L^2} \right)^2 \right]
\ee
This form of the near-horizon geometry has manifest $SO(2,1)\times G\times U(1)$ symmetry as guaranteed by the theorem proved in~\cite{FKLR}. For our solution we have
\bea
\Gamma(x)= \frac{\beta}{4} (\xi+x^2), \qquad A(x)^2=\epsilon(1-x^2), \qquad B(x)^2= \frac{\epsilon P(x)}{(1-x^2)^{n-1}(\xi+x^2)}
\eea
with $A_0$ given by (\ref{A0}), $\kappa$ given by (\ref{xidef}) and $P(x)$ by (\ref{P}). This solution has the scaling symmetry $(\beta, v) \to (K \beta, K^{-1}v)$ where $K>0$. This allows one to set the constant $\beta$ to any desired value (i.e. it is a redundant parameter). It is worth pointing out that if one analytically continues the AdS$_2 \to S^2$ we have an Einstein metric which falls in the general class derived in~\cite{Mann2006,HOY}.

\subsubsection{Summary of solutions}
We have derived a set of near-horizon geometries of the form~(\ref{gNH}), and satisfying $R_{\mu\nu} = \Lambda g_{\mu\nu}$ in $D=2n+2$ dimensions for $n\geq 2$,  given by
\bea\label{localmet}
&&\gamma_{AB}dx^Adx^B = \epsilon L^2 \left[  \frac{(\xi+x^2)(1-x^2)^{n-1} dx^2}{P(x)} +\frac{P(x) }{ (\xi+x^2)  (1-x^2)^{n-1}} (d\phi+\sigma)^2 \right. \nonumber \\ && \left. \qquad \qquad \qquad \qquad + (1-x^2) \bar{g}_{ab} d\bar{x}^a d\bar{x}^b \right] \\
&& h_Adx^A =  \pm \frac{2\epsilon \sqrt{\xi} P(x) }{(\xi+x^2)^2 (1-x^2)^{n-1}} (d\phi+\sigma) - \frac{2x}{\xi+x^2} dx \\
&&F= \frac{-2n\epsilon +\lambda(1+\xi)}{L^2(\xi+x^2)} + \frac{4\xi P(x)}{L^2 (1-x^2)^{n-1}(\xi+x^2)^3}
\eea
where $P(x)$ is a polynomial given by (\ref{P}) and $\lambda \equiv \Lambda L^2$. The metric $\bar{g}_{ab}$ is any $(2n-2)$-dimensional K\"ahler-Einstein metric such that $\bar{R}_{ab}=2n\bar{g}_{ab}$ and $J=d\sigma/2$ is the K\"ahler form.  For fixed choice of the K\"ahler-Einstein structure $(\bar{g},J)$, the solution is parameterised by the constants $(\xi, c, L)$ where $ \xi, L >0$, $c$ is the integration constant occurring in $P(x)$ and $\epsilon = \pm 1$. Note that the parameter $\beta$ has cancelled from the solutions as it is redundant (see above). Therefore this constitutes a three parameter family of metrics.  It is worth noting that although they are valid for $n\geq 2$,  if one sets $n=1$ and formally $\bar{g}_{ab}=0$, $\sigma=0$, then the solution is locally isometric to the near-horizon limit of extremal Kerr-NUT-(AdS$_4$)~\cite{KL1}. These solutions may thus be regarded as a generalisation to all even dimensions. It is worth pointing out that these solutions are valid for any $\Lambda$.

\section{Global analysis of horizon geometries}
In this section we will derive the conditions necessary for extending our local horizon metric to a complete Riemannian metric on a compact smooth manifold $\mathcal{H}$ with no boundary (i.e. a cross section of the horizon). We will only analyse the case $\Lambda \leq 0$. First note that since we require $\gamma_{AB}$ to be a positive definite metric we must have $A^2 \geq 0$ (with a possible equality only at isolated points) and therefore either $\epsilon=1$ and $x^2 \leq 1$ or $\epsilon=-1$ with $x^2 \geq 1$. For $\epsilon=+1$ we must have $P(x) \geq 0$, whereas for $\epsilon=-1$ we must have $(-1)^n P(x) \geq 0$, in other words $\epsilon^n P(x) \geq 0$.

We first consider potential singularities in the metric as one varies $x$. Inspecting the metric we see that these can only occur at $x=\pm 1$, $x=\pm \infty $ or the roots of $P(x)$. Following the terminology of~\cite{PP} we refer to these points as endpoints. A complete manifold requires $x_1 \leq x \leq x_2$ where $x_1<x_2$ are two adjacent endpoints and that the singularities at these endpoints  are removable by coordinate transformations. Compactness further requires that $x_1,x_2$ are finite endpoints (i.e the metric distance between points in $x_1<x<x_2$ and the endpoints is finite). This leaves a number of possibilities: the endpoints can be either at $\pm 1$ or at simple zeros of $P(x)$. Note that regularity of the metric requires that if either $x_1$ or $x_2$ are equal to $\pm 1$ then $P(x)$ must also vanish at these points (otherwise, for example, the norm of $\partial /\partial \phi$ diverges at these endpoints). Therefore in all cases we must have $x_1 \leq x \leq x_2$ with $P(x_1)=P(x_2)=0$ and $\epsilon^nP(x)>0$ for $x_1<x<x_2$. Note that this implies $\epsilon^nP'(x_1)>0$ and $\epsilon^nP'(x_2)<0$.  For later use it is convenient to note the identity
\be
\label{Pder}
P'(x_i)=\frac{(\xi+x_i^2)(1-x_i^2)^{n-1}[\lambda\epsilon (1-x_i^2)-2n]}{x_i}
\ee
which can be derived using (\ref{Peq}) and $P(x_i)=0$. It is worth noting that necessarily $x_i \neq 0$, since all roots of $P$ are non-zero\footnote{This follows from the fact that $P(0)=(2n-\lambda \epsilon)\xi$. For $\epsilon=1$ we see that $P(0)>0$. For $\epsilon=-1$ we see that $P(0)\neq 0$ unless $\lambda=-2n$, and thus for simplicity we assume $\lambda \neq -2n$.}.

The possibilities for the endpoints are listed in the following table:
\begin{table}[!h]
\label{table}
\centering
\begin{tabular}{ | c | c | c | c | c | c | c | c | c | }
\hline
Case & I & IIa & IIb & IIIa & IIIb & IV & Va & Vb\\
\hline
$x_1$ & $-1$ & $-1$ & $>-1$ & $<-1$  & $1$ & $>-1$ & $>1$ & $<-1$\\
$x_2$ & $1$ & $<1$ & $1$   & $-1$   & $>1$  & $<1$ & $>1$ & $<-1$ \\
 \hline
 \end{tabular}
 \caption{Endpoints}
\end{table}
%\vskip.1cm

\noindent Note that in case IIa (case IIb), the inequality $x_2<1$ ($x_1>-1 $) follows by the assumptions that $x_2\neq 1$ ($x_1 \neq 1$) and $x_2$ ($x_1$) is an adjacent endpoint to $-1$ ($1$).  Also note that case IIa and IIb, case IIIa and IIIb, and case Va and Vb, can be mapped into each other using the freedom in the definition of $x \to -x$. Therefore without loss of generality we refer to these cases as II, III, V respectively, and we need only consider the five cases I-V.

In fact, case III and V cannot occur. This is easy to see as follows. Simply note that in both cases $\epsilon=-1$ and thus without loss of generality  one has a root $x_2>1$ such that $\textrm{sgn}\; P'(x_2) = (-1)^{n+1}$ (i.e. case IIIb and case Va). However equation (\ref{Pder}) implies $\textrm{sgn}\; P'(x_2) = (-1)^{n}$ and therefore we have a contradiction. In Appendix \ref{caseII} we show case II cannot be made smooth with compact topology. This leaves case I and case IV, and as we will show these can be made smooth with compact topology, and end up homeomorphic to $S^{2n}$ and $S^2$-bundles over compact K\"ahler-Einstein manifolds $K$, respectively.

Before moving on we note that the horizon metric has conical singularities at the roots $x_i$ of $P(x)$ as long as $x_i^2 \neq 1$. Removal of the conical singularity at $x=x_i$ is equivalent to
\be
\label{period}
\Delta \phi = 4\pi (\xi+x_i^2) \frac{\epsilon^{n-1}(1-x_i^2)^{n-1}}{| P'(x_i) |} = 4\pi \left| \frac{x_i}{\lambda\epsilon(1-x_i^2) -2n} \right|
\ee
where the second equality follows from the identity (\ref{Pder}).

\subsection{Inhomogeneous $S^{2n}$}
In this section we analyse case I listed in table 1. Note that since in this case $-1 \leq x \leq 1$ we must have $\epsilon=1$.
From the explicit expression for $P(x)$ given in equation (\ref{P}) we see that $P(1)=0=P(-1)$ implies $c=0$ and therefore $P(x)$ is an even function. The constraint $P(1)=0$ now provides one linear equation for the parameter $\xi$.  One can solve this by performing the various binomial sums involved, to get an explicit value for $\xi$ given by
\be
\label{xiMP}
\xi= \xi_* \equiv \frac{2n+1-\lambda}{(2n+1)(2n-1-\lambda)} \; .
\ee
Substituting this value back into $P(x)$ and simplifying gives, after some work,\footnote{In fact it is easier to go back and solve (\ref{Peq}).}
\be
P(x)= (1-x^2)^n  \left[ 1+\xi_* - \frac{\lambda (1-x^2)}{2n+1} \right]  \; .
\ee
It is worth noting that for $\lambda=0$ we have $\xi_*=1/(2n-1)$, and the solution is much easier to obtain using the identity $Q_n(u,1/(2n-1))= (1-u)^n/(2n-1)$.
Putting all this together gives the following horizon metric
\bea
&&\gamma_{AB}dx^Adx^B= L^2 \left[ \frac{(\xi_*+x^2) dx^2}{(1-x^2) \left[ 1+\xi_* - \frac{\lambda (1-x^2)}{2n+1}  \right] } + \frac{(1-x^2) \left[ 1+\xi_* - \frac{\lambda (1-x^2)}{2n+1}  \right] }{\xi_*+x^2} (d\phi+\sigma)^2 \right. \nonumber \\  && \qquad \qquad  \qquad \qquad \left. +(1-x^2) \bar{g}_{ab}d\bar{x}^ad\bar{x}^b \right]
\eea
where $\xi_*$ is given by (\ref{xiMP}).

The above metric is smooth and invertible for $-1<x<1$. We must now check regularity at $x=\pm 1$. Set $x=\pm (1- \eta^2)$ and expanding for small $\eta$ one gets
\be
\gamma_{AB}dx^Adx^B = 2L^2 \left[ d\eta^2+ \eta^2 ( (d\phi+\sigma)^2 +\bar{g}_{ab}d\bar{x}^ad\bar{x}^b ) \right] + \dots
\ee
where $\dots$ signify terms higher order in the $\eta$ expansion.  Smoothness at $\eta=0$ requires that $\Delta \phi=2\pi$ and that $\bar{g}$ be the Fubini-Study metric on $K=\mathbb{CP}^{n-1}$. The horizon metric then looks like the origin of $\mathbb{R}^{2n}$ near $x= \pm 1$. The horizon topology in this case is then $\mathcal{H}=S^{2n}$.

To summarise, this case gives a 1-parameter (given by $L$) family of inhomogeneous horizon geometries on $\mathcal{H}=S^{2n}$. In fact these near-horizon geometries are isometric to the near-horizon limits of the extremal Myers-Perry-(AdS) in $2n+2$ dimensions with all angular momenta equal.  For $\lambda=0$  these near-horizon limits were calculated in~\cite{FKLR} and it is easy to check that they are the same by setting $x=\cos\theta$ and
\be
\label{MPparam}
L^2= \frac{2na^2}{2n-1} \; .
\ee
It can be checked that our solution for $\Lambda<0$ is isometric to the near-horizon geometry of extremal Myers-Perry-AdS in $2n+2$ dimensions with all angular momenta equal, although we will not give the details here.

\subsection{Inhomogeneous $S^2$ bundles over K\"ahler-Einstein spaces} \label{analysis}
In this section we analyse case IV listed in table 1.
This corresponds to the generic case when the endpoints are $-1<x_1<x_2<1$ and thus $\epsilon=1$. This implies that $A^2 =(1-x^2)>0$ for all $x_1\leq x \leq x_2$ and $P(x)>0$ for $x_1<x<x_2$.  It follows that we must have $P'(x_1)>0$ and $P'(x_2)<0$. The horizon metric is thus smooth and invertible for $x_1<x<x_2$ with potential conical singularities at $x=x_1,x_2$.  From (\ref{period}) we see that simultaneous removal of these singularities implies
\be
-\frac{\lambda(1-x_1^2)-2n}{x_1} = \frac{\lambda (1-x_2^2)-2n}{x_2} \; .
\ee
For $\lambda=0$ this immediately implies that $x_1=-x_2$. For $\lambda<0$ notice that $f(x)=[\lambda(1-x^2)-2n]/x$ is a monotonically increasing function for all $x>0$. Therefore the regularity condition $f(x_2)=-f(x_1)$ implies $x_1=-x_2$. From the form of $P(x)$ the condition $x_1=-x_2$ implies $c=0$. The period of $\phi$ is  given by (\ref{period}) and is simply
\be
\label{deltaphi}
\Delta \phi= \frac{4\pi x_2}{2n-\lambda(1-x_2^2)} \; .
\ee
This means that for a fixed point on the base $K$ the $(x,\phi)$ part of the metric is smooth and of $S^2$ topology. Compactness of $\mathcal{H}$ then clearly requires $K$ to be compact.

So far we have only considered regularity of the horizon metrics when one varies $x$. In this case (in contrast to the other cases considered) we have not constrained the K\"ahler-Einstein space $(K,J,\bar{g})$. We must also impose that $d\phi+\sigma$ is independent of the coordinate chart on $K$ used. This implies that $\int _C 2J$ over any 2-cycle $C$ in $K$ must be an integer multiple of $\Delta \phi$, and hence $\int_{C} 2J$ and $\int_{C'} 2J$ for any two 2-cycles $C,C'$ must be rationally related. This constraint is automatically satisfied for K\"ahler-Einstein manifolds since the first Chern class of its tangent bundle, which is an integral class, is given by $c_1(K)= [ \bar{\rho}/2\pi] = (n/2\pi) [2J]$ ($\bar{\rho}$ is the Ricci form of $\bar{g}$).  Let $p$ be the Fano index of the K\"ahler-Einstein base $K$, which by definition is the largest positive integer such that $p^{-1}c_1(K)$ is an integral class. Now consider a set of 2-cycles  $\Sigma_i \subset K$ which form a representative basis of the free part of $H_2(K,\mathbb{Z})$. It follows that
\be
\label{c1K}
\int_{\Sigma_i} c_1(K) = n_i p
\ee
for a set of integers $n_i \in \mathbb{Z}$ such that gcd$(n_i)=1$; note that without loss of generality we can always take $n_i$ to be non-negative. It follows that
\be
\label{2Ji}
\int_{\Sigma_i} 2J = n_i \frac{2\pi p}{n}
\ee
and thus for any integral 2-cycle $C=\sum_i c_i \Sigma_i$ we have
\be
\left| \int_C 2J \right| = \left|\sum_i c_i n_i \right|\frac{2\pi p}{n}  \geq \frac{2\pi p}{n}
\ee
where the last equality follows from the fact one can always find integers $c_i$ such that $\sum_i c_i n_i=1$\footnote{This is a basic generalisation of a number theory result called B\'ezout's identity.}. It follows that the minimum absolute value of $\int_C 2J$ over all possible 2-cycles is then simply $2\pi p /n$. Therefore the period of $\phi$ must satisfy
\be
\label{deltaphim}
m \Delta \phi = \frac{2\pi p}{n}
\ee
for some positive integer $m$.  In fact for any compact K\"ahler-Einstein manifold $p \leq n$ with equality if and only if $K=\mathbb{CP}^{n-1}$.

Equating the two expressions (\ref{deltaphi}) and (\ref{deltaphim}) for $\Delta \phi$ thus gives the following quantisation condition
\be
\label{quant}
m= \frac{p[2n-\lambda(1-x_2^2)]} {2n x_2}  \; .
\ee
Note that this is particularly simple for $\lambda=0$ which gives $x_2=p/m$.

The existence of smooth metrics in this case thus boils down to proving that the (even) polynomial $P(x)=2nQ_n(x^2,\xi)-\lambda Q_{n+1}(x^2,\xi)$ must have a smallest positive root $x_2<1$, such that $x_2$ satisfies  (\ref{quant}) for some integer $m$. If this can be achieved then the horizon metric is a smooth inhomogeneous metric on a compact manifold which is a fibre bundle (possibly trivial) over $K$ with $S^2$ fibre.

To prove the existence of $x_2$ first note the identities:
\be
P(0)= (2n-\lambda)\xi, \qquad P(1)= \frac{n! \sqrt{\pi}(2n-1-\lambda)(\xi-\xi_*)}{\Gamma(n+\frac{1}{2})}
\ee
where $\xi_*$ is given by (\ref{xiMP}). For $\lambda \leq 0$ it is clear that $P(0)>0$. Furthermore if $\xi<\xi_*$ then $P(1)<0$. It follows that, for $\lambda\leq 0$, if $\xi<\xi_*$ then there must exist a root $x_0$ of $P(x)$ in the interval $0<x<1$. It remains to show that $x_0$ is the smallest positive root, i.e. $P(x)>0$ for $-x_0<x<x_0$, so that $x_2=x_0$. Fortunately this is easy to prove. Suppose that $x_0$ is not the smallest positive root. It follows that there must exist a root $x_- \leq x_0<1$ such that $P'(x_-)\geq 0$. However from the identity (\ref{Pder}) we see that any root $0<x_i<1$ must have $P'(x_i)<0$. This is a contradiction and hence $x_0$ must be the smallest positive root as required. Therefore we have proved that a sufficient condition for the existence of $x_2$ is $\xi<\xi_*$.  We can also prove the condition $\xi<\xi_*$ is necessary as follows. First note that $\xi=\xi_*$ was analysed earlier and has $x_2=1$ and thus we discard in this case. Now assume $\xi>\xi_*$ so that $P(1)>0$. Therefore, either $P(x)>0$ for all $0\leq x \leq 1$ or there exists a root $0<x_-<1$ such that $P'(x_-) \geq 0$. As argued above this is a contradiction and therefore we have shown that if $\xi\geq \xi_*$ then $P(x)>0$ for all $0<x<1$, i.e. $x_2$ does not exist.
To summarise, in the $\lambda\leq 0$ case we have shown that a root $0<x_2<1$, such that $P(x)>0$ for $x^2\leq x_2^2$, exists if and only if $\xi<\xi_*$.

Finally it remains to show that the quantisation condition (\ref{quant}) can always be satisfied for some positive integer $m$. This can be established as follows.  First note that for $\lambda \leq 0$ the function $[2n-\lambda (1-x^2)]/x$ is monotonically decreasing, and takes the value $+\infty$ at $x=0$ and $2n$ at $x=1$. Therefore (\ref{quant}) has solutions if and only if $m$ is any integer satisfying
\be
\label{mbound}
m> p \; .
\ee
It is worth noting that one can solve (\ref{quant}) explicitly for $x_2=x_2(m,p)$ (e.g. for $\lambda=0$ it is just $x_2=p/m$) which in turn gives $\xi=\xi(m,p)$ via the identity,
\be
\xi= -\frac{x_2 \left[ C_{2n-1}^{(-n+\frac{1}{2})}(x_2)- \frac{\lambda}{2n+1} C_{2n+1}^{(-n-\frac{1}{2})}(x_2) \right]}{ C_{2n-2}^{(-n+\frac{1}{2})}(x_2) - \frac{\lambda}{2n+1} C_{2n}^{-n-\frac{1}{2}}(x_2)}
\ee
which follows from $P(x_2)=0$ and (\ref{QGegen}).

In summary,  we have smooth horizon metrics which are $S^2$-bundles over a compact K\"ahler-Einstein space $K$ if and only if $\xi<\xi_*$ and (\ref{quant}) are satisfied.  For a given base $K$ (which gives $p$), these metrics are parameterised by one continuous parameter $L>0$ and an integer $m$ satisfying the bound (\ref{mbound}).

\subsubsection{The topology of $\mathcal{H}$ }

The topology  of our bundles depends on the integers $m,p$ and thus we will refer to these horizons by $\mathcal{H}_{m,p}$. In Appendix~\ref{topH} we derive some of the basic invariants of these spaces, which we summarise at the end of this section.  Recall though that the two topological restrictions we are interested in are:  is $\mathcal{H}_{m,p}$ cobordant to $S^{2n}$? Is $\mathcal{H}_{m,p}$ positive Yamabe type? Fortunately these are easy to deal with as follows.

First, observe that any $S^2$-bundle over an oriented compact manifold is oriented cobordant to a sphere. This can be seen as follows. Let $S^2 \to \mathcal{H} \to K$ be any $S^2$-bundle over a compact manifold $K$, with dim $\mathcal{H}=N$. The structure group of such bundles is $SO(3)$. One can construct an associated ball bundle over $K$. That is, one replaces the fibres $S^2$ with the $3$-ball $B_3$ (so $\partial B_3 \cong S^2$) and then constructs the associated bundle using the same $SO(3)$ transition functions acting on $B_3 \subset \mathbb{R}^3$. Call this new $N+1$ dimensional (oriented) manifold $X$ so that by construction $\partial X =\mathcal{H}$. It is then clear that a (oriented) cobordism between $\mathcal{H}$ and $S^{N}$ exists. One simply cuts a sufficiently small $(N+1)$-ball $B_{N+1}$ from the interior of $X$ so that the remaining manifold has boundaries $\mathcal{H}$ and $S^N$ as required. Explicitly, the cobordism is given by $\Sigma=X \backslash B_{N+1}$. Therefore we immediately deduce that our $\mathcal{H}_{m,p}$ are always (oriented) cobordant to $S^{2n}$ for any $m>p$ and $n\geq 2$.

Next, observe that any $S^2$-bundle over a compact manifold $K$, such that the base admits a positive Ricci curvature metric, must also admit a positive Ricci curvature metric~\cite{Nash}. Since  by assumption our $K$ is compact and Einstein with positive scalar curvature, we immediately deduce that our $\mathcal{H}_{m,p}$ must admit a positive Ricci curvature metric. It follows that $\mathcal{H}_{m,p}$ are all of positive Yamabe type for any $m>p$ and $n \geq 2$. \\

\noindent {\it Summary of topology: } We will now summarise the key topological properties of  the $2n$-dimensional manifolds $\mathcal{H}_{m,p}$ where $m,p$ are positive integers such that $m>p$.
\begin{itemize}
\item  For $n=2$ it is either homeomorphic to the trivial bundle $S^2\times S^2$ (if $m$ even) or the non-trivial $S^2$-bundle over $S^2$ which is $\mathbb{CP}^2\# \overline{\mathbb{CP}^2}$ (if $m$ is odd). For $n \geq 3$ it is always a non-trivial $S^2$-bundle over a compact positive K\"ahler-Einstein manifold $K$ where $p$ is its Fano index; furthermore different $m$ give different topologies.
\item They are simply connected for all $n \geq 2$.
\item They are spin if and only if $m+p$ is even, for all $n \geq 2$.
\item The Euler characteristic is $2\chi(K)$ for any $n\geq 2$ and all $m$.
\item They are oriented cobordant to $S^{2n}$ for all $n\geq 2$ and all $m$.
\item They are positive Yamabe type for all $n \geq 2$ and all $m$.
\end{itemize}
The derivation of the $n=2$ case is discussed in the next section and the rest are discussed in detail in  Appendix~\ref{topH}.

\subsubsection{Four dimensional horizons: $S^2 \times S^2$ and $\mathbb{CP}^2 \# \overline{\mathbb{CP}}^2$}
In this section we consider the $n=2$ case explicitly. We note that the only K\"ahler-Einstein space in this case is $S^2$ with (note the normalisations)
\be
\bar{g}_{ab}d\bar{x}^a d\bar{x}^b = \frac{1}{4}\left[ d\theta^2+\sin^2\theta d\chi^2 \right], \qquad \sigma= \frac{\cos\theta}{2} d\chi
\ee
and $\chi\sim \chi+2\pi$. Since there is only one 2-cycle, $S^2$ itself, the invariant $\int_{S^2} 2J=2\pi$, i.e. $p=2$. Therefore, for every integer $m>2$ we have an explicit smooth inhomogeneous metric on $S^2$ bundles over $S^2$.  As is well known~\cite{Stenrod}, up to homeomorphism, there are only two types of $S^2$-bundles over $S^2$ since they are classified by $\pi_1(SO(3))=\mathbb{Z}_2$. One is the trivial bundle $S^2\times S^2$ and the other is a non-trivial bundle which has the same topology as
 $\mathbb{CP}^2 \# \overline{\mathbb{CP}^2}$. In our case these arise depending on whether $m$ is even or odd respectively. This fact can be deduced from our analysis as follows (see~\cite{GMSW1} for a simple argument). Since $p=2$ we note that $\mathcal{H}_{m,2}$ is spin if and only if $m$ is even.  Since $S^2\times S^2$ is spin and $\mathbb{CP}^2 \# \overline{\mathbb{CP}^2}$ is not spin, it immediately follows that $m$ even corresponds to the former and $m$ odd to the latter.

It is worth pointing out that the $\lambda=0$ solution is particularly simple. Then
\be
\label{Pneq2}
P(x)= \frac{4}{3}x^4+4(\xi-1)x^2+4\xi  \; .
\ee
The existence of a positive root $x_2<1$ occurs if and only if $\xi<1/3$  and explicitly is given by
\be
x_2^2= \frac{3}{2} \left[ 1-\xi- \sqrt{ (1-\xi)^2-\frac{4\xi}{3}} \right]  \; .
\ee
Since we must have $\xi>0$ we see that $0<x_2<1$ is uniquely parameterised by $0<\xi<1/3$ (the function $x_2(\xi)$ is a monotonically increasing function on the domain $[0,1/3]$ with range $[0,1]$).  The quantisation condition is simply  $ x_2= \frac{2}{m}$, which allows one to give a simple explicit expression for $\xi$:
\be
\xi= \xi_m \equiv \frac{4}{3}\left( \frac{ 3- \frac{4}{m^2} }{4+m^2} \right)  \label{xim}
\ee
where recall the integer $m>2$.  Note that in this parameterisation, the quartic~(\ref{Pneq2}) is simply:
\begin{equation}
P(x) = (4-m^2x^2)\left(\xi_m - \frac{4 x^2}{3m^2}\right) \; .
\end{equation}
In section \ref{summary} we give the explicit form of the full near-horizon geometry in this simple case.

\subsubsection{Examples in all even dimensions}
We now discuss some interesting examples in dimensions $n>2$. \\

\noindent{$K=\mathbb{CP}^{n-1}$}: \\
Here we consider the K\"ahler base to be $K=\mathbb{CP}^{n-1}$ which has a unique Einstein metric given by the Fubini-Study metric, and this has $p=n$. It is a homogeneous metric with $SU(n)$ isometry. Therefore, the horizon and near-horizon geometries are both cohomogeneity-1. The resultant near-horizon geometries then posses an isometry group $SO(2,1)\times SU(n)\times U(1)$.

This class is the natural generalisation of the $n=2$ case. Recall that for $n=2$ we showed $\mathcal{H}$ is homeomorphic to the trivial bundle $S^2\times S^2$ (if $m$ is even) or the non-trivial bundle $\mathbb{CP}^2 \# \overline{\mathbb{CP}^2}$ (if $m$ is odd). However, for $n\geq 3$ the possible topologies of our solutions are very different to the $n=2$ case, despite the similarly of the local forms of the metric. In particular our $\mathcal{H}_{m,p}$ is never homeomorphic to the trivial bundle or $ \mathbb{CP}^n\# \overline{\mathbb{CP}^n}$. To see this observe that for $n \geq 3$, different $m$ must give different topologies (as we show in the Appendix \ref{topH}), and $m=0$ and $m=1$ correspond to the trivial bundle and $\mathbb{CP}^n\# \overline{\mathbb{CP}^n}$ respectively~\cite{PP}. However regularity of our bundles requires $m>n$, and thus these two topologies are immediately ruled out as claimed.\\

\noindent {$\textrm{dim} \, \mathcal{H}=6$}: \\
This corresponds to the case $n=3$, which corresponds to $D=8$ dimensional near-horizon geometries. The K\"ahler-Einstein space $K$ in this case is 4 dimensional and spatial sections of the horizon are six dimensional $S^2$-bundles over $K$. In fact, all K\"ahler-Einstein metrics on complex 2-manifolds $(K,J)$ with positive curvature have been classified~\cite{Tian}.  These occur exactly on $\mathbb{CP}^2$, $\mathbb{CP}^1 \times \mathbb{CP}^1$, or the del Pezzo surfaces $dP_k= \mathbb{CP}^2 \# k \overline{\mathbb{CP}^2}$ for $3\leq k \leq 8$ (i.e. $\mathbb{CP}^2$ blown up at $k$ points in a general position), and the K\"ahler-Einstein metric is uniquely determined by $(K,J)$.  Each of these provides us with a near-horizon geometry of the form derived earlier.

Let us now consider the isometry groups of these K\"ahler-Einstein metrics. In fact  $\mathbb{CP}^2$, $\mathbb{CP}^1 \times \mathbb{CP}^1$ and $dP_3$ are all toric manifolds (i.e. admit an effective $U(1)^2$-action), and their associated K\"ahler-Einstein metrics\footnote{Note that this metric on $dP_3$ is only known numerically~\cite{HW}. } must be invariant under $U(1)^2$. Further, the $\mathbb{CP}^2$ and $\mathbb{CP}^1\times \mathbb{CP}^1$ cases have enhanced symmetry of $SU(3)$ and $SU(2)^2$ respectively.  Recall our corresponding near-horizon geometries have isometry groups $SO(2,1)\times U(1)\times G$ where $G$ is the isometry group of $K$, and thus the cases $\mathbb{CP}^2$ and $\mathbb{CP}^1\times \mathbb{CP}^1$ are both cohomogeneity-1.  On the other hand, $dP_k$ for $4 \leq k \leq 8$ are not toric manifolds and their K\"ahler-Einstein metrics generically have no continuous isometries. Therefore, choosing $K= dP_k$ for $4 \leq k \leq 8$ gives us examples of near-horizon geometries with only $U(1)$ rotational isometry, or in total $SO(2,1) \times U(1)$.  Interestingly, for $k \geq 5$ these solutions possess $2k-8$ extra continuous parameters, corresponding to the complex structure moduli of $dP_k$ (these correspond to the positions of the $k$ blow up points). \\

\noindent {$\textrm{dim} \, \mathcal{H} \geq 8$}: \\
This corresponds  to $n>3$ and is qualitatively similar to the $n=3$ case just described, although there is no analogous classification of possible compact positive curvature K\"ahler-Einstein manifolds available yet.  We can give some explicit examples though.

If we choose the K\"ahler-Einstein space to be homogeneous then the horizon, and near-horizon geometry, is cohomogeneity-1. The most symmetric case is $K=\mathbb{CP}^{n-1}$ which we have already considered above.  Another homogeneous possibility is to take $K={(\mathbb{CP}^1)}^{\times (n-1)} \cong {(S^2)}^{\times (n-1)}$. The resultant near-horizon geometry then possess an isometry group $SO(2,1)\times SO(3)^{\times (n-1)} \times U(1)$.

More generally consider the case when $K$ is a toric manifold so the isometry group is $U(1)^{n-1}$.  It follows that the rotational isometry group of the associated near-horizon geometries is $U(1)^n$. These horizon and near-horizon geometries are generically cohomogeneity-$n$. A special case of course includes the homogeneous examples above.

Finally the most extreme case occurs for K\"ahler-Einstein metrics that possess no symmetries whatsoever (generalising the higher del Pezzo surfaces). In this case the $S^2$ bundle over $K$ has only one $U(1)$ symmetry and the associated near-horizon geometry has isometry $SO(2,1) \times U(1)$.

\section{Area and angular momentum formulas}
In this section we will discuss various physical properties of the near-horizon geometries we have derived.  Recall that, for any compact K\"ahler-Einstein manifold $K$, we found a family of compact horizon geometries parameterised by one positive real number $L$ (which sets the scale of the horizon) and an integer $m \geq p$ where $p$ is the Fano index of $K$. Our horizons $\mathcal{H}_{m,p}$ have their topology  determined by the integers $(m,p)$. For $m=p$ one must have $K=\mathbb{CP}^{n-1}$ and so $p=n$, and the horizons have $S^{2n}$ topology and are isometric to those of extremal Myers-Perry with all angular momenta equal. For $m>p$ the topology of the horizon is of an $S^2$-bundle over any $K$.
Recall the coordinate ranges are $-x_2 \leq x \leq x_2$ and $\phi \sim \phi+ \Delta\phi$,  where $x_2$ is the smallest positive root of $P(x)$ and depends on the integer $m$. Note that $\Delta \phi= \frac{2\pi p}{nm}$ for $m>p$ and $\Delta \phi= 2\pi$ for $m=p$. We first give the volume form on the horizon
\be
\epsilon = \sqrt{\gamma} \; dx^1 \wedge \cdots \wedge dx^{2n}=  L^{2n} (1-x^2)^{n-1} dx \wedge d\phi \wedge \bar{\epsilon}
\ee
where we have chosen an orientation and $\bar{\epsilon}$ is the volume form associated to the K\"ahler-Einstein space $K$.

\subsection{Area}
The area of a cross section of the horizon is
\be
A(\mathcal{H}_{m,p})= \int_{\mathcal{H}} \epsilon= - L^{2n} \Delta \phi \;  \textrm{vol}(K) \frac{(n-1)! \sqrt{\pi}}{\Gamma(n+\frac{1}{2}) } C^{(-n+\frac{1}{2})}_{2n-1}(x_2)
\ee
where $\textrm{vol}(K)= \int_K \bar{\epsilon}$ and we have used
\be
\sum_{l=0}^{n-1}  {}^{n-1}C_l  \; \frac{(-1)^l 2 x^{2l+1}}{2l+1} = - \frac{(n-1)! \sqrt{\pi}}{\Gamma(n+\frac{1}{2}) } C^{(-n+\frac{1}{2})}_{2n-1}(x)
\ee
where $C^{(\alpha)}_n(x)$ is a Gegenbauer polynomial.

 The $\mathcal{H}=S^{2n}$ case, $\xi=\xi_*$, corresponds to $x_2=1$, $\Delta \phi=2\pi$,  $\textrm{vol}(\mathbb{CP}^{n-1})= \pi^{n-1}/(n-1)!$, and using $C^{(-n+\frac{1}{2})}_{2n-1}(1)=-1$, gives\footnote{The area of a unit round $S^n$ is given by $A_n=2 \pi^{\frac{n+1}{2}}/ \Gamma( \frac{n+1}{2})$ (note that there is a typo in~\cite{FKLR}). The volume of $\mathbb{CP}^{n-1}$ with the Fubini-Study metric normalised as in this paper, can be deduced from the volume of an $S^{2n-1}$ with unit round metric, $A_{2n-1}$, using the fact it can be written as a Hopf fibration. Explicitly $A_{2n-1}= \Delta \phi \; \textrm{vol}(\mathbb{CP}^{n-1})$. }
\be
A(S^{2n})= \frac{2\pi^{n+\frac{1}{2}}L^{2n} }{\Gamma(n+\frac{1}{2})} = A_{2n} L^{2n}
\ee
where $A_{2n}$ is the volume of a unit round sphere. Note that for $\lambda=0$ this agrees with the Myers-Perry value~\cite{FKLR} upon using (\ref{MPparam}) (as it should).

\subsection{Angular momentum}
The Komar angular momentum of the near-horizon geometry is given by~\cite{FKLR}
\be
\label{Jkomar}
J_i \equiv J[m_i]= \frac{1}{16 \pi G} \int_\mathcal{H} \sqrt{\gamma}\;  h \cdot m_i
\ee
where $m_i$ is a rotational Killing field. For our near-horizon geometry the available rotational Killing fields are
\be
\label{Hkilling}
m_\phi \equiv \frac{\Delta \phi}{2\pi} \frac{\partial}{ \partial \phi},  \qquad m_i =\bar{m}_i
\ee for $1 \leq i \leq d-1$ where $d \leq n$ depends on $K$,  where $\bar{m}_i$ are the commuting Killing fields of $K$.  Note that if $K$ is toric then $d=n$. We have defined $m_\phi$ such that its orbits are canonically normalised with period $2\pi$.

The angular momentum associated to $m_\phi$ evaluates to
\be
J_{\phi}= \pm \frac{L^{2n} (\Delta \phi)^2 \; \textrm{vol}(K) \sqrt{\xi}}{ 32 \pi^2 G} \int_{-x_2}^{x_2} \frac{2P(x)}{(\xi+x^2)^2} dx   \; .
\ee
The integral for $J_{\phi}$ can be done by parts using (\ref{Peq}), resulting in
\be
\int_{-x_2}^{x_2} \frac{2P(x)}{(\xi+x^2)^2} dx=  -\frac{2}{x_2} \left[ 2n Q_{n-1}(x_2^2, -1)-\lambda Q_{n}(x^2_2, -1)  \right]
\ee
where $Q_n(u,\xi)$ is the polynomial (\ref{Qpol}). For the case at hand
\be
Q_{n}(x^2, -1)= \sum_{l=0}^n \frac{{}^n C_l (-1)^l x^{2l}}{2l-1}=- \frac{n! \Gamma(\frac{3}{2})}{\Gamma(n+\frac{3}{2} )} C_{2n}^{(-n-\frac{1}{2})}(x)
\ee
where $C^{(\alpha)}_n(x)$ is a Gegenbauer polynomial.  Putting all this together gives
\be
\label{Jphi}
J_\phi = \pm \frac{L^{2n} (\Delta \phi)^2 \; \textrm{vol}(K) A_{2n}}{16 \pi^2 G A_{2n-1}} \frac{n\sqrt{\xi}}{x_2} \left[ C_{2n-2}^{(-n+\frac{1}{2})}(x_2) -\frac{\lambda}{2n+1} C_{2n}^{(-n-\frac{1}{2})}(x_2) \right]
\ee
where $A_n$ is the area of a unit round $n-$sphere as above.

In the case when $K=\mathbb{CP}^{n-1}$ we have checked explicitly, in Appendix \ref{IntJ}, that the angular momenta associated to the internal Killing fields $\bar{m}_i$ of $\mathbb{CP}^{n-1}$ are
\be
\bar{J}_i=0
\ee
for $1 \leq i \leq n-1$. It also follows that these angular momenta vanish for direct products of lower dimensional complex projective spaces. In fact in Appendix \ref{IntJ} we also show that for general toric $K$ these internal angular momenta must vanish as well (one does not need the explicit metric for this calculation).

For the $\mathcal{H}=S^{2n}$ case the formulas for $J_{\phi}$ can be simplified using
%$Q_{n}(1,-1)=-n! \sqrt{\pi}/\Gamma(n+\frac{1}{2})$ (or
$C_{2n}^{(-n-\frac{1}{2})}(1)=2n+1$, resulting in
\be
J_\phi = \pm  \frac{A_{2n} L^{2n}}{8\pi G }  \sqrt{\xi_*} n \left(2n-1 -\lambda \right)  \; .
\ee
When $\lambda=0$ this gives $J_{\phi}= \pm A_{2n} L^{2n} n \sqrt{2n-1}/(8\pi G)$ which agrees with~\cite{FKLR} upon using (\ref{MPparam}).

\subsection{Area versus angular momentum curves}
Let us first consider the simplest case of $\mathcal{H}= S^{2n}$  corresponding to the near-horizon geometry of an extremal Myers-Perry(-AdS) black hole. For $\lambda=0$, using the formula for the area of the horizon and the angular momentum we find
\be
\label{MParea}
A(S^{2n})= \frac{8\pi G}{n \sqrt{2n-1}} | J_{\phi} |  \; .
\ee
For $\lambda<0$ one can also write $J_\phi$ as a function of the horizon area:
\be
|J_\phi| = \frac{nA(S^{2n})}{8 \pi G } \sqrt{\frac{\left[ 2n-\Lambda \left(\frac{A(S^{2n})}{A_{2n}} \right)^{1/n}\right]^2-1}{2n+1}}  \; .
\ee
From this expression it is easy to show that, for fixed $J_\phi$, $A(S^{2n})$ decreases monotonically as one makes $\Lambda$ more negative starting from $\Lambda=0$. This makes intuitive sense, as turning on a negative cosmological constant makes gravity more attractive.

These area versus angular momentum curves can also be written for our new near-horizon geometry. For $\lambda=0$ it reads:
\be
A(\mathcal{H}_{m,p})= \frac{8\pi G}{x_2 \sqrt{\xi} } \left[ -\frac{x_2 C^{(-n+\frac{1}{2})}_{2n-1}(x_2)}{C^{(-n+\frac{1}{2})}_{2n-2}(x_2)} \right] |J_\phi |  = \frac{8 \pi G   \sqrt{\xi}}{x_2} |J_\phi|
\ee
where the second equality follows from the identity:
\be
\xi= -\frac{x_2 C_{2n-1}^{(-n+\frac{1}{2})}(x_2)}{C_{2n-2}^{(-n+\frac{1}{2})}(x_2)}
\ee
which is a consequence of $P(x_2)=0$ and (\ref{QGegen}).  For $\lambda<0$ one can write down an analogous curve for $\mathcal{H}_{m,p}$ although it is not so revealing.

We can compare the curve for our new near-horizon geometry to the spherical topology case. For $\lambda=0$ we find that, at fixed $J_\phi$,
\be
\frac{A(\mathcal{H}_{m,p})}{A(S^{2n})} =  \frac{n}{x_2} \sqrt{ \xi (2n-1) } = n \sqrt{ -\frac{(2n-1) C_{2n-1}^{(-n+\frac{1}{2})}(x_2)}{x_2 C_{2n-2}^{(-n+\frac{1}{2})}(x_2)}} .  \label{genratio}
\ee
We have explicitly checked that (\ref{genratio}) is a monotonically decreasing function of $x_2$ in the interval $0<x_2<1$ for low values of $n$. Assuming this is the case for all $n$ (as seems reasonable), it follows that
\be
n< \frac{A(\mathcal{H}_{m,p})}{A(S^{2n})} <n \sqrt{2n-1}  \; .
\ee
Thus in particular, for fixed $J_\phi$, the area of $\mathcal{H}_{m,p}$ is always larger than that of the spherical topology case. It is worth emphasising that if there are new black hole solutions corresponding to our near-horizon geometries, the canonical rotational Killing field $m_\phi$ need not correspond to the same combination of rotational Killing fields at asymptotic infinity as it does for the Myers-Perry solution (see next section). Therefore, the above comparison of the area at fixed $J_\phi$ may not be meaningful outside the context of this class of near-horizon geometries.

\section{Which horizon geometries arise from new black holes?}
\label{BH}
In this section we will investigate the possibility that the $D=2n+2 \geq 6$ dimensional near-horizon geometries we have found are in fact the near-horizon limits of yet to be known, stationary extremal black hole solutions to $R_{\mu\nu}=\Lambda g_{\mu\nu}$ for $\Lambda \leq 0$. We will focus on asymptotically flat black holes ($\Lambda=0$) and asymptotically globally AdS black holes ($\Lambda<0$)\footnote{We remark that our near-horizon geometries can be obtained as ``near-horizon" limits of spacetimes with Taub-NUT like asymptotics. These spacetimes can be obtained by analytically continuing certain Einstein metrics in~\cite{HOY}. However, due to their asymptotics, they necessarily have closed time like curves everywhere.}.  Near spatial infinity, asymptotically these black hole spacetimes would look like
\bea
&&ds^2 \sim -\left(1-\frac{\Lambda R^2}{D-1} \right)dt^2 + \frac{dR^2}{1-\frac{\Lambda R^2}{D-1}} +R^2 ds^2(S^{2n})  \label{asymp}\\
&& \quad ds^2(S^{2n})= \sum_{I=1}^n d\mu_I^2 +\mu_I^2 (d\psi^I)^2 \label{evensphere}
\eea
as $R\to \infty$, where $R$ is some radial coordinate, $\sum_{I=1}^n \mu_I^2=1$ are the latitude coordinates on $S^{2n}$ and $\psi^I \sim \psi^I +2\pi$. Note that the Killing fields $\psi_I=\partial / \partial \psi^I$ are the standard generators of the Cartan subgroup $U(1)^n \subset SO(2n+1)$.  There are a number of constraints on the symmetries and topologies of the horizons of such black holes which we now recall.

In both of these cases, the maximal rotational isometry group is $SO(2n+1)$ and its maximal abelian subgroup is $U(1)^n$. Furthermore, for a rotating black hole, the rigidity theorem~\cite{HIW,IM} guarantees the existence of at least one $U(1)$ isometry (although this has only been proved for non-extremal black holes\footnote{See \cite{HIRig} for partial results on the extremal case.}). Therefore black holes of this kind have an isometry group whose abelian subgroup is $\mathbb{R}\times U(1)^d$ such that $1\leq d \leq n$ (as shown above though, asymptotically they do have the maximal abelian symmetry $\mathbb{R}\times U(1)^n$).

Our near-horizon geometries are $D=2n+2$ dimensional spacetimes which satisfy $R_{\mu\nu}=\Lambda g_{\mu\nu}$ for $\Lambda \leq 0$. Their isometry groups are $SO(2,1)\times G\times U(1)$ where $G$ is the isometry group of $K$. The $SO(2,1)$ component is typical for near-horizon geometries and is guaranteed by general theorems regarding symmetry enhancement~\cite{KLR, FKLR}. The isometry group of spatial sections of the horizon $\mathcal{H}$ is (by assumption) $G\times U(1)$ and thus depends on the choice of $K$. If $K$ is toric then the maximal abelian subgroup of $G$ is $U(1)^{n-1}$ and thus in this case the total abelian isometry group of the near-horizon geometries is $\mathbb{R}\times U(1)^n$. If $K$ has a metric with no isometries then the total abelian isometry group of the near-horizon geometry is $\mathbb{R}\times U(1)$.

For asymptotically flat, or globally AdS black holes the exterior to the black hole defines a cobordism from $\mathcal{H}$ to $S^{2n}$.  As discussed earlier all our horizons geometries are guaranteed to be cobordant to a sphere since they are $S^2$-bundles over a compact manifold.  Furthermore for asymptotically flat black holes the horizon must be positive Yamabe type, which is also the case for our horizon geometries.

Therefore in all even dimensions greater than four we have found examples of near-horizon geometries such that the spatial sections of the horizon are not of spherical topology, but still consistent with all known symmetry and topology constraints required for asymptotically flat and globally AdS black holes. It is therefore natural to speculate whether these are the near-horizon geometries of yet to be found extremal black holes in such spacetimes. We now expand on some examples in more detail.

\subsection{Black holes with $\mathbb{R}\times U(1)^{[(D-1)/2]}$ symmetry}
For this discussion we will focus on the class of near-horizon geometries with the K\"ahler-Einstein base $K=\mathbb{CP}^{n-1}$ for $ n\geq 2$. Recall that for $n=2$ then $\mathcal{H}$ is either the trivial bundle $S^2\times S^2$ (for $m$ even) or the non-trivial $S^2$-bundle over $S^2$ (for $m$ odd) which is homeomorphic to $\mathbb{CP}^2\# \overline{\mathbb{CP}}^2$. For $n \geq 3$ then $\mathcal{H}$ is always a non trivial $S^2$-bundle over $\mathbb{CP}^{n-1}$ and for different $m$ they have different topology (in this case $m>n$).

The rotational symmetry of these near-horizon geometries is $SU(n)\times U(1)$.  As discussed above the abelian subgroup of this is $U(1)^n$, the maximal abelian rotational symmetry group possible for asymptotically flat or globally AdS black holes in $2n+2$ dimensions. These isometries are associated to angular momenta $J_i$ for $i=1, \dots , n$.  We calculated these with respect to the Killing fields on the horizon (\ref{Hkilling}) and found $J_\phi \neq 0$ (\ref{Jphi}), and that the ``internal" angular momenta $\bar{J}_i=0$ for $1\leq i \leq n-1$ (see Appendix \ref{IntJ}).

It is natural to expect that the putative black hole solutions would have the same rotational symmetry as the horizon geometry, and thus a total of $\mathbb{R}\times SU(n)\times U(1)$. For example, the $\mathcal{H}=S^{2n}$ case we derived earlier which also has $SU(n)\times U(1)$ rotational symmetry, arises from the extremal Myers-Perry-(AdS) black hole with equal angular momenta which does have the enhanced isometry group $\mathbb{R}\times SU(n)\times U(1)$. However, for our new horizon geometries $\mathcal{H}_{m,n}$ will in fact now argue that this cannot be the case.  Suppose that the full black hole solution does have a global $SU(n)\times U(1)$ isometry with orbits which are $U(1)$ bundles over $\mathbb{CP}^{n-1}$. This gives a natural way of identifying the rotational Killing fields on the horizon (\ref{Hkilling}),  in terms of those of the total rotational symmetry $SO(2n+1)$ at infinity.  Simply write the round $S^{2n}$ in (\ref{asymp}) in terms of a round $S^{2n-1}$, and in turn, the round $S^{2n-1}$ as a Hopf fibration over $\mathbb{CP}^{n-1}$, i.e.
\be
ds^2(S^{2n}) = d\theta^2+ \sin^2\theta[(d\phi'+\sigma')^2+\bar{g}']
\ee
where $\bar{g}'$ is the Fubini-Study metric (normalised by $\textrm{Ric}(\bar{g}')=2n \bar{g}'$) with K\"ahler form $J'=d\sigma'/2$ and of course $\Delta \phi'=2\pi$.  It is then easy to show that in terms of the standard set of rotational Killing fields $\psi_I$ for $1\leq I \leq n$ of $S^{2n}$ defined in (\ref{evensphere}),  we have
\be
\frac{\partial}{\partial \phi'} = \sum_{I=1}^n \psi_I, \qquad \bar{m}'_i= \psi_n-\psi_i \ee
for $1\leq i \leq n-1$, where $\bar{m}'_i$ are the $U(1)^{n-1}$ generators on $\mathbb{CP}^{n-1}$.  Now, due to the assumption of the global $SU(n)\times U(1)$ isometry, it is most natural to identify the data $(\phi', \sigma', \bar{g}')$ in this sphere at infinity with the corresponding data on the horizon ($\phi, \sigma, \bar{g}$) (since we are choosing $K=\mathbb{CP}^{n-1}$). However, since $\Delta \phi= 2\pi /m \neq \Delta \phi'$, and $(\bar{g}, J)$ and $(\bar{g}', J')$ are both normalised in the same way, we find a contradiction. Therefore this argument implies, surprisingly, that there can be no asymptotically flat or globally AdS black hole with a global $SU(n)\times U(1)$ rotational isometry and horizon geometry $\mathcal{H}_{m,n}$. This is unfortunate as it means we can say less about any potential black hole solutions with such horizons. For example, the above identification would have allowed one to deduce the angular momenta as viewed from infinity are all equal. Also, the problem of determining such black hole solutions would have been more tractable, since the black hole metrics would have been cohomogeneity-2 and thus one could cast the problem on the 2 dimensional orbit space $\hat{M}=M/[\mathbb{R}\times SU(n)\times U(1)]$ (analgous to the Weyl solutions case).

In the absence of any other symmetry, there is no natural way to identify the data on the horizon $(\phi, \sigma, \bar{g})$ with the data at infinity $(\phi, \sigma', \bar{g}')$. However, note that one has the same number of commuting Killing fields on the horizon and the sphere at infinity (this is true for any toric $K$). Therefore, a natural expectation for the symmetries of the hypothetical black hole solutions is that they have the same commuting rotational symmetries as the near-horizon geometry, i.e. a global isometry group $\mathbb{R}\times U(1)^n$.  This then allows one to identify the commuting rotational Killing fields in the most general possible way: namely, the two sets of commuting Killing fields $(\partial_\phi, \bar{m}_i)$ and $(\partial_{\phi'}, \bar{m}_i')$ are necessarily related by some constant matrix in $SL(n,\mathbb{Z})$. Since we do not know the explicit form of this constant matrix, we cannot deduce what the angular momenta would be from the point of view of asymptotic infinity; in particular, they need not have all angular momenta equal.

Finally, it is worth noting that in the generic toric $K$ case, $\mathcal{H}_{m,p}$ has isometry group $U(1)^n$ and thus one might expect any corresponding full black hole solutions to also have $\mathbb{R}\times U(1)^n$ symmetry. This would then allow one to identify the commuting Killing fields on the horizon and at infinity via some constant matrix, as in the $K=\mathbb{CP}^{n-1}$ case just discussed.

\subsection{Black holes with $\mathbb{R}\times U(1)$ symmetry}\label{onerot}
In this section we consider the class of near-horizon geometries we have derived where the K\"ahler-Einstein base $K$ is chosen so that it has no isometries at all. Such spaces are known to exist in four and higher dimensions.  Therefore the associated near-horizon geometries in this case start in eight dimensions.  The near-horizon geometries in this case would have total isometry group $SO(2,1)\times U(1)$. If they arose as near-horizon limits of an extremal black hole, it would necessarily have only $\mathbb{R}\times U(1)$ symmetry. In this case the near-horizon geometry only has one angular momentum, corresponding to the $U(1)$ generator $\partial / \partial \phi$, and as we showed this is necessarily non-zero. Since the horizon has less abelian symmetry than asymptotic infinity there is no natural way to identify the $U(1)$ generators and therefore we cannot guess what angular momenta these black holes would carry relative to infinity.

Let us now mention some explicit examples. The simplest case for which we can have such symmetry is $n=3$, i.e. 4d K\"ahler-Einstein space $K$. The only non-toric possibilities are the complex del Pezzo surfaces $dP_k$ for $4\leq k \leq 8$.  Therefore  we have established the existence of examples of horizon geometries which are fully consistent with the black hole horizon topology theorems, but also have only one $U(1)$ rotational symmetry. As mentioned earlier, the $dP_k$ for $k\geq 5$ have a moduli space of complex structures of dimension $2k-8$, corresponding to the freedom in placing the blowup points. Hence the associated near-horizon geometries possess these additional continuous parameters. These parameters could in turn constitute further continuous `hair' for the putative black holes.

For $n>3$ the K\"ahler-Einstein space $K$ is $2n-2\geq 6$ dimensional. Much less is known about the classification of such spaces. However, presumably it is the case there are examples with no isometries, as well as continuous moduli.

Therefore, in all even dimensions $D\geq 8$ we have candidate near-horizon geometries for new asymptotically flat or globally AdS extremal black holes with just $\mathbb{R}\times U(1)$ isometry.

\subsection{Associated (boosted) black strings}
In this section we point out that given our Ricci-flat near-horizon geometries in $2n+2$ dimensions with horizon section $\mathcal{H}$, we may trivially construct Ricci-flat near-horizon geometries in $2n+3$ dimensions with horizon section $S^1\times \mathcal{H}$.

This construction is analogous to that used in~\cite{FKLR} where, for every extremal Myers-Perry black hole in $2n+2$ dimensions, a boosted extremal black string was constructed in $2n+3$ dimensions. Its near-horizon geometry then provides an example of one with $\mathcal{H}=S^1\times S^{2n}$ and $U(1)^{n+1}$ rotational isometry. This is the correct maximal abelian rotational symmetry for an asymptotically flat black hole in $2n+3$ dimensions. This led~\cite{FKLR} to conjecture that a subset of these solutions could be the near-horizon limit of yet to be found extremal asymptotically flat black rings in odd dimensions (in 5d one can explicitly check this is true as solutions are known~\cite{KLR}).

 In the present case, obviously we do not have the full even dimensional black hole solution (i.e. the analogue of the Myers-Perry solution with our horizon $\mathcal{H}_{m,p}$) or for that matter know whether it actually exists. Nevertheless, one may still construct the near-horizon geometry of the corresponding boosted black string as follows. Consider the direct product of our Ricci flat solutions with a line $dz^2$. Now consider the ``boost"
\be
\phi \to \phi + s_\beta \Omega z \qquad  z \to  c_\beta \; z
\ee
where $s_\beta\equiv \sinh \beta$, $c_\beta \equiv \cosh\beta$ and $\Omega$ are constants. The resultant near-horizon geometry is
\bea
&&ds^2=\Gamma(x) [A_0 r^2 dv^2+2dvdr ] \\
&& \qquad + L^2 \left[  A(x)^2 \bar{g}_{ab} d\bar{x}^a d\bar{x}^b+ \frac{dx^2}{B(x)^2} + B(x)^2 \left(d\phi+\sigma+ s_\beta \Omega dz + \frac{\kappa  rdv}{L^2} \right)^2 \right] + c_\beta^2 dz^2 \nonumber
\eea
which is Ricci flat in $2n+3$ dimensions. One can perform a global analysis of the horizon metrics and the result are smooth compact horizon geometries with topology of $\mathcal{H}_{m,p}
 \times S^1$, where $\mathcal{H}_{m,p}$ is identical to the $2n$ dimensional horizons we have derived and $z$ must be periodic on the $S^1$. Thus we have a family with four continuous parameters $(L,\beta, \Omega, \Delta z)$ and an integer $m>p$, although it is expected that $\Omega$ would be related to $L$ if a black hole exists (see below).  As a consequence of our construction, the isometry group of these near-horizon geometries is $SO(2,1)\times G\times U(1)^2$. Note that $S^1\times \mathcal{H}_{m,p}$ is cobordant to $S^{2n+1}$ and positive Yamabe type\footnote{In fact $S^1\times M^n$ is cobordant to $S^{n+1}$ for any closed manifold $M^n$. This is simply because it is the boundary of the $n+2$ dimensional manifold $D\times M^n$ where $D$ is a disk. Also $S^1\times M^n$ is  positive Yamabe type if $M^n$ is.}. For $K=\mathbb{CP}^{1}$ for example, we have $D=7$ smooth cohomogeneity-1 near-horizon geometries with $\mathcal{H}=S^1 \times S^2 \times S^2$,  abelian symmetry $U(1)^{2}$, which are cobordant to $S^{5}$ and positive Yamabe type.

The constants $\beta$ and $\Omega$ introduced above have the following significance. Suppose that there exists an asymptotically flat extremal black hole with a horizon geometry given by our $\mathcal{H}_{m,p}$ and consider the associated black string obtained by adding a line $dz^2$. Now boost $(t,z) \to ( c_\beta t-s_\beta z, c_\beta z -s_\beta t)$ where $t$ is the asymptotically flat time coordinate of the lower dimensional black hole. Upon taking the near-horizon limit of this black string one will end up with a near-horizon geometry that is related to that of the black hole as above where $\Omega$ is the angular velocity of the black hole in the $\phi$ direction (it is not possible to calculate $\Omega$ form knowledge of the near-horizon geometry alone~\cite{FKLR}).

Note that these hypothetical black strings would have standard KK asymptotics (i.e. $\mathbb{R}^{1,2n+1}\times S^1$).  Generically such strings possess a tension. Typically one can choose the boost $\beta=\beta_{crit}$ such the tension vanishes. In this case, one might expect new asymptotically flat black holes in $\mathbb{R}^{1,2n+2}$ to exist with horizon topology $S^1\times \mathcal{H}_{m,p}$. Physically, this is achievable because the straight string can be ``bent'' into a ring in such a way as to have Minkowski asymptotics while inputting no energy. But since near-horizon geometries are independent of the asymptotic geometry, the near-horizon geometry of strings with $\beta=\beta_{crit}$ is expected to be the same as such higher dimensional ``black rings". As mentioned above, this is precisely what occurs for extremal black rings and the Kerr black string in $D=5$ \cite{KLR}. It is natural to expect a similar phenomenon to occur here and so we have candidate near-horizon geometries for asymptotically flat extremal black holes in $2n+3$ dimensions with horizon topology $S^1\times \mathcal{H}_{m,p}$.

\section{Discussion}
In this paper, we have investigated the space of allowed extremal black hole horizon topologies, for vacuum general relativity, including a negative cosmological constant, in even dimensions greater than four. Our strategy was based on the observation that for such black holes, Einstein's equations can be decoupled and solved on the horizon $\mathcal{H}$ alone. We have constructed an infinite class of (non-static) vacuum near-horizon geometries in $D=2n+2$ for $n\geq 2$ (see section \ref{summary} for the $n=2$ Ricci flat case and~(\ref{localmet}) for the general local form of the solutions). A global analysis of the local horizon metrics reveals that the topology of $\mathcal{H}$ is either an $S^2$-bundle over any compact K\"ahler-Einstein manifold $K$, or $S^{2n}$. Regularity implies that the $S^2$-bundle topology solutions are characterised, for a given base $K$,  by a continuous parameter (essentially the angular momentum) and a positive integer $m$ satisfying some bound determined by the topology of $K$. The spherical topology solutions are simply the near-horizon geometries of extremal Myers-Perry-(AdS) with all angular momenta equal. We emphasise that all our new near-horizon geometries have horizons of \emph{non}-spherical topology (with the precise topology determined by the choice of $K$ and the integer $m$).

 First, consider the simplest case $n=2$. Then for our new near-horizon solutions, $\mathcal{H}$ must have topology $S^2 \times S^2$ or $\mathbb{CP}^2 \#\overline{\mathbb{CP}}^2$.  We have analogous near-horizon geometries in higher dimensions $n>2$, where $K$ can be any $(2n-2)$-dimensional K\"ahler-Einstein space. In contrast to the $n=2$ case, the horizon $\mathcal{H}$ is always a non-trivial $S^2$-bundle over $K$, and furthermore, for a fixed base $K$, one has an infinite number (countable) of distinct horizon topologies.
As explained in the Introduction, the horizon topology of asymptotically flat or globally AdS black holes is constrained by the existence of a cobordism from $\mathcal{H}$ to a sphere, and the generalisation of Hawking's topology theorem (positive Yamabe). In fact all our horizon topologies automatically satisfy these restrictions. We emphasise that all these conclusions are equally valid with and without a negative cosmological constant. Therefore, our work raises the possibility that there exist extremal, asymptotically flat or globally AdS black holes with precisely these non-spherical horizon geometries.

Our work leaves a number of open questions. Most interestingly, of course, is whether there are in fact asymptotically flat or AdS extremal black holes with near-horizon limits given by our new near-horizon geometries.  If this is the case then is reasonable to expect non-extremal and non-vacuum generalisations as well. These black holes would have to be rotating and thus constructing them would be a difficult task.

An important feature of our analysis is that it yields, in addition to the topology, the explicit horizon geometries for the proposed extremal black holes. Although asymptotic information is lost in the near-horizon limit (e.g. angular velocities, mass), we can still compute the conserved angular momenta and area from the near-horizon geometry alone. In fact for our near-horizon geometries we find only one independent Komar angular momentum $J$. Interestingly, in the pure vacuum case ($\Lambda = 0$) we find that for fixed $J$, the area of the non-spherical horizons is always \emph{more} than that of the corresponding Myers-Perry horizons. However, for  the corresponding candidate non-spherical extremal black holes, we cannot say how this angular momentum $J$ would be distributed at asymptotic infinity, and thus this comparison should be taken with caution.

%Hence from standard arguments we infer that these candidate non-spherical extremal black holes are entropically favoured within an ensemble defined by fixing total near-horizon angular momentum (usually one fixes the mass but this data is not available from the near-horizon geometry).

Another interesting point is that for fixed angular momentum $J$, we can have more than one near-horizon solution. In particular, in $D=6$ for fixed $J$, we can have an infinite number of near-horizon geometries with $\mathcal{H}=S^2\times S^2$ (even $m$) but also an infinite number of near-horizon geometries with $\mathcal{H}=\mathbb{CP}^2 \# \overline{\mathbb{CP}^2}$ (odd $m$). Thus we see two phenomena: discrete hair for a fixed topology, and non-uniqueness of the near-horizon geometry for fixed $J$. On the other hand, in higher (even) dimensions $D\geq 8$, we do not have the discrete hair for fixed topology, although we do have infinite non-uniqueness of the near-horizon geometry for fixed $J$ (since we have an infinite number of distinct possible horizon topologies with the same $J$). It should be emphasised though, that since we do not have the full black hole solutions available (or even know they exist), we can not compute the mass and thus we cannot say anything about the corresponding black hole (non-)uniqueness problem.

One particularly interesting example of a near-horizon geometry arises when we choose the K\"ahler-Einstein base to have no isometries (e.g. the del Pezzo surfaces $dP_k$ for $4 \leq k \leq 8$).  The resulting near-horizon geometry then has just $SO(2,1)\times U(1)$ symmetry. If it arises as the near-horizon limit of an extremal black hole, the black hole would have at most $\mathbb{R}\times U(1)$ symmetry. This is the minimum symmetry requirement for a rotating black hole (all known solutions have more symmetries), and in fact it has been conjectured that black hole solutions with this symmetry should exist~\cite{Reall}. It is worth noting that evidence for asymptotically flat rotating black holes with this minimal symmetry has been found recently using a variety of different approaches~\cite{EHNO, DFMRS}. In particular, the analysis of linearized gravitational perturbations of \emph{odd} dimensional $D \geq 9$ Myers-Perry black holes with equal angular momenta suggests a new branch of solutions with $\mathbb{R}\times U(1)$ symmetry~\cite{DFMRS}\footnote{It is worth noting that previously~\cite{KLRm1}, for the odd dimensional $D \geq 7$ Myers-Perry-AdS with equal angular momenta, an instability was found (which sets in for sufficiently large rotation and includes the extremal limit), and was conjectured to have an endpoint which is a non-rotating black hole with even less symmetry, i.e. just stationary but not axisymmetric.}. Such black holes have an upper bound on their angular momenta which is saturated when they are extremal, and the instability sets in for angular momenta larger than some lower bound, which of course includes the extremal limit. However, this potential new branch of solutions would still have spherical horizon topology.

More approachable open problems involve the analysis of near-horizon geometries. In this paper we found a class of solutions within some general ansatz (i.e. near-horizons of the form (\ref{gamma}) and (\ref{h})). An interesting question is whether we have in fact found all solutions within this ansatz or not. We note that one can in fact classify compact Einstein spaces of the same form as our horizon geometries and the possible topologies one finds are $S^{2n}$, $\mathbb{CP}^n$ and a finite discrete family of $S^2$-bundles over K\"ahler-Einstein base $K$ of the same form as ours (although we have an infinite discrete family for fixed $K$)~\cite{PP}. It is thus striking that a regular $\mathcal{H}=\mathbb{CP}^n$ case does not arise in our solutions (we did find an example with a conical singularity though). Therefore it is possible that there are other solutions within this class which would give such horizon topologies. We note that for $n$ odd $\mathbb{CP}^n$ is cobordant to  $S^{2n}$ and therefore consistent with all known restrictions on asymptotically flat or AdS black holes.

In this paper we have focussed our attention on the classification problem of higher-dimensional black holes. Since we included a negative cosmological constant, our results have potential applications to AdS/CFT. Most obviously, the existence of non-spherical horizon topology black holes which are asymptotic to global AdS is currently unknown. These would correspond to some interesting phase of the dual gauge theory on $\mathbb{R}\times S^3$. Various attempts at finding AdS black rings have so far failed\footnote{Although see~\cite{CER} for some approximate constructions. See also~\cite{KL1} for an approximate near-horizon geometry.}. In particular the existence of supersymmetric AdS$_5$ black rings with $\mathbb{R}\times U(1)^2$ symmetry has been ruled out~\cite{KLR0,KL0}, suggesting that the known spherical horizon topology solutions~\cite{CCLP, KLR-1} may in fact be the most general ones. However in this paper we have found non-supersymmetric non-spherical topology (and non-black ring like) examples. For instance, in 6d we have explicit $S^2\times S^2$ topology extremal horizon metrics. It is thus natural to wonder whether these are in fact near-horizon geometries of yet to be found extremal AdS black holes. If so, then it seems reasonable non-extremal generalisations should exist.  Such objects would correspond to novel thermal phases of the dual gauge theory. \\

\par \noindent \emph{Acknowledgements} -  We would like to thank Jerome Gauntlett, Don Page, James Sparks,  Dan Waldram and Toby Wiseman for useful conversations.  We would especially like to thank Jerome Gauntlett for comments on a draft of our paper, and James Sparks for valuable correspondence on topology.  HK is supported by a fellowship from the Pacific Institute of Mathematical Sciences and NSERC. JL is supported by an EPSRC career acceleration fellowship.

\appendix

\section{Local properties of horizon metrics}
We consider $2n$-dimensional Riemannian manifolds  $(\mathcal{H}, \gamma)$ (spatial sections of the horizon) endowed with metrics of the form:
\be
\gamma_{AB}dx^Adx^B=L^2\left[ d\rho^2+ A(\rho)^2 \bar{g}_{ab}d\bar{x}^a d\bar{x}^b + B(\rho)^2
(d\phi+\sigma)^2  \right]
\ee where $\bar{g}$ is a K\"ahler metric on the $2n-2$ dimensional K\"ahler-Einstein `base' space $K$ with K\"ahler form $\bar{J} = \frac{1}{2}d\sigma$ and curvature normalised such that $\textrm{Ric}(\bar{g})=2n\bar{g}$. Lower Latin indices $a,b, c,\dots$ take values from $1, \cdots, 2n-2$. Henceforth we adopt the convention all quantities defined on the base are ``barred", unless otherwise stated. Note that $L$ is a parameter with dimensions length. We introduce a vielbein basis $\{ e^A: A \in \{ 0, a, 2n-1 \}\}$ defined by
\be
e^0=L d\rho, \qquad e^a=L A \bar{e}^a, \qquad e^{2n-1}=L B(d\phi+A)
\ee with $\bar{e}^a$ a veilbein basis for $(K,\bar{g})$. Note we also sometimes use the coordinate $dx=B d\rho$.
\subsection{Curvature calculations}\label{Curvature}
In this subsection we denote $\rho$-derivatives by $d/d\rho =\; '$. The spin connection, defined by $de^A=-\omega^A_{\phantom{A}B} \wedge e^B$, is readily computed:
\begin{eqnarray}
&&\omega_{0a}= -\frac{A'}{LA} e^a, \qquad \omega_{0 \, 2n-1}= -\frac{B'}{L B}
e^{2n-1},
\qquad \omega_{ab}=\bar{\omega}_{ab}- \frac{B}{L A^2} \bar{J}_{ab} e^{2n-1},  \nonumber \\
&&\omega_{a\, 2n-1}=-\frac{B}{L A^2} \bar{J}_{ab} e^b \; .
\end{eqnarray} The curvature two-form $\Theta_{AB} = d\omega_{AB} + \omega_{AC}\wedge \omega^{C}_{\phantom{C}B}$ has the following non-vanishing components:
\bea
&&\Theta_{0a}=\frac{1}{L^2}\left[- \frac{A''}{A} e^0 \wedge e^a +\frac{B}{A^2}\left(
\frac{A'}{A}-\frac{B'}{B} \right) e^{2n-1}\wedge e^b \bar{J}_{ab} \right]  \\
&&\Theta_{0 \, 2n-1}= \frac{1}{L^2}\left[- \frac{B''}{B} e^0 \wedge e^{2n-1} +\frac{B}{A^2}\left(
\frac{A'}{A}-\frac{B'}{B} \right) e^{a}\wedge e^b \bar{J}_{ab} \right]  \\
&&\Theta_{a \, 2n-1} = \frac{1}{L^2}\left[\frac{B}{A^2}\left(\frac{A'}{A}-\frac{B'}{B} \right)
\bar{J}_{ab} e^0 \wedge e^b + \left( \frac{B^2}{A^4} -
\frac{A'B'}{AB}\right) \delta_{ac} e^c \wedge e^{2n-1} \right]  \\
&& \Theta_{ab}= \frac{1}{L^2}\left[\frac{2B}{A^2} \left( \frac{A'}{A}-\frac{B'}{B}
\right) e^0 \wedge e^{2n-1} \bar{J}_{ab} + e^c \wedge e^d \left( -\frac{A'^2}{A^2}
\delta^a_c\delta^b_d - \frac{B^2}{A^4} \bar{J}_{ac}\bar{J}_{bd} - \frac{B^2}{A^4}
\bar{J}_{ab}\bar{J}_{cd} \right) \right] \nonumber \\ &&\qquad \quad+ \; \bar{\Theta}_{ab} \; .
\eea Note that we have used the fact $\bar{\nabla}\bar{J}=0$ and $\bar{J}^{a}_{\phantom{a}b}\bar{J}^{b}_{\phantom{b}c} = -\delta^{a}_{\phantom{a}c}$ to simplify the above expression. Finally, the Riemann curvature is read off from $\Theta_{AB} = \frac{1}{2}R_{ABCD}e^C \wedge e^D$. A calculation yields the following independent components (for clarity we will label the $2n-1$ basis component simply as $\phi$ below):
\begin{eqnarray}
R_{0a0b} &=& -\frac{A''}{L^2 A}\delta_{ab}, \qquad R_{0a\phi b} = \frac{B}{L^2A^2}\left(\frac{A'}{A} - \frac{B'}{B}\right)\bar{J}_{ab}, \qquad R_{0\phi 0 \phi} = -\frac{B''}{L^2 B} \nonumber \\
R_{0\phi a b} &=& \frac{2 B}{L^2 A^2}\left(\frac{A'}{A} - \frac{B'}{B}\right)\bar{J}_{ab}, \qquad R_{a\phi b\phi} = \left(\frac{B^2}{A^4}-\frac{A'B'}{AB}\right)\frac{\delta_{ab}}{L^2}, \\
R_{abcd} &=& \frac{1}{L^2}\left[\frac{\bar{R}_{abcd}}{A^2} + 2\left(-\frac{A'^2}{A^2}\delta_{a[c}\delta_{d]b} + \frac{B^2}{A^4}\left(\bar{J}_{a[c}\bar{J}_{d]b} - \bar{J}_{ab}\bar{J}_{cd}\right)\right)\right] \nonumber
\end{eqnarray} From here it is straightforward to calculate the Ricci tensor~(\ref{Ricci}).

\subsection{Conformal K\"ahler structure}\label{appendix:kahler}
In this subsection we denote $x$-derivatives by $d/dx = \; '$. Define the two-forms
\be
\label{Jpm}
J_{\pm} \equiv  L^2 A^2 \bar{J} \pm e^0 \wedge e^{2n-1}= L^2[A^2 \bar{J}\pm dx\wedge (d\phi+\sigma)]
\ee
which satisfy ${J_{\pm}}^{A}_{\; B} {J_{\pm}}^B_{\; C}=-\delta^A_B$ (i.e. they are almost complex structures). One can show that
\be
dJ_{\pm}= \left( \frac{2A'}{A}\mp \frac{2}{A^2} \right) dx \wedge J_{\pm}  \; .
\ee
Since $\bar{J}$ is the K\"ahler form on the base $K$, the volume form $\bar{\epsilon}=\bar{J}^{n-1}/(n-1)!$ and thus $\bar{J}^n=0$. It follows that
\be
J_{\pm}^n= \pm n (L^2A^2)^{n-1} \bar{J}^{n-1} \wedge e^0 \wedge e^{2n-1}=  \pm n! \; \epsilon
\ee
where $\epsilon$ is the volume form of $(\mathcal{H},\gamma)$.
Now, since $K$ is an $n-1$-dimensional complex manifold there is an $(n-1,0)$ form $\Omega$ on $K$ such that $\bar{J}^{n-1} = i^{n-1}(-1)^{(n-1)(n-2)/2}\Omega \wedge \bar{\Omega}$ where $\bar{\Omega}$ is the complex conjugate of $\Omega$. This allows one to write $J_{\pm}^n=i^n(-1)^{n(n-1)/2} \Omega_{\pm} \wedge \bar{\Omega}_{\pm}$ where
\be
\label{Omegapm}
\Omega_{\pm}=\sqrt{ \frac{n}{2}} (L^2A^2)^{\frac{n-1}{2}} \; \Omega \wedge \left(e^0 \pm i  e^{2n-1} \right)   \; .
\ee
One can check that
\be
d\Omega_{\pm} =  \frac{(A^{n-1}B)'}{A^{n-1}B} dx\wedge \Omega_{\pm} \; .
\ee
From these local properties it follows that $\mathcal{H}$ is a complex manifold with complex structure specified by $\Omega_{\pm}$, and  furthermore $(\mathcal{H}, \gamma)$ is conformally K\"ahler (see~\cite{GMSW} for a similar argument).

In fact from (\ref{Omegapm}) we may read off a set of complex coordinates for $\mathcal{H}$. First let $z^i$ be a set of complex coordinates for $K$. Then note that $\sigma=-i(\partial-\bar{\partial})k$ where $k$ is the K\"ahler potential of $K$. However since $\partial k$ is a $(1,0)$-form on $K$ we must have $\Omega\wedge \partial k=0$. This allows us to write
\be
\label{Omegadw}
\Omega_{\pm}= \sqrt{ \frac{n}{2}} (L^2A^2)^{\frac{n-1}{2}} B \; \Omega \wedge dw^{\pm}
\ee
where
\be
w^{\pm} = \pm [i \phi-k(z^i,\bar{z}^i ) ]+ \int^x \frac{dx}{B^2} \; .
\ee
Thus $Z^I=(w^{\pm},z^i)$ are two different sets of complex coordinates on $\mathcal{H}$.

Now let us consider our explicit solution. Using $A^2=(1-x^2)$ we get
\be
dJ_{\pm}= \mp \frac{2}{1\mp x} dx \wedge J_{\pm}  \;
\ee
and therefore
\be
\hat{J}_{\pm} =  \frac{J_{\pm}}{(1\mp x)^2}
\ee
is closed. The 2-form $\hat{J}_{\pm}$ is a K\"ahler form and the associated K\"ahler metric is
\be
\hat{\gamma}^{\pm}_{AB}=\frac{\gamma_{AB}}{(1\mp x)^{2}}  \; .
\ee It is useful to have an expression for the Ricci form of this K\"ahler metric. Recall that in general for a K\"ahler metric $g_{\mu\bar{\nu}}$, in some set of complex coordinates $z^{\mu}$, the Ricci form is
\be
\rho= -i \partial \bar{\partial} \log {\det (g_{\mu\bar{\nu}})} = db, \qquad b=  \frac{i}{2}(\partial -\bar{\partial}) \log {\det ( g_{\mu\bar{\nu}} )}
\ee
where the expression in terms of $b$ follows from the identity $\partial\bar{\partial} = -\frac{1}{2}d(\partial-\bar{\partial})$.
 Therefore one simply needs $\det{g_{\mu\bar{\nu}}}$ in a set of complex coordinates. For the case at hand, using (\ref{Omegadw}) one can calculate $\epsilon$ (and thus $\det{\gamma}$ from $\epsilon=\pm i^n \det (\gamma_{Z^I\bar{Z}^J} )\; dw^{\pm} \wedge d\bar{w}^{\pm} \wedge dz^1\wedge d\bar{z}^1 \dots \wedge dz^{n-1}\wedge d\bar{z}^{n-1}$) in the coordinates $Z^I=(w^{\pm}, z^i)$. Taking care of the conformal rescaling
it follows that
\be
{ \det (\hat{\gamma}^{\pm}_{Z^I \bar{Z}^j})} =  \frac{(L^2 A^2)^{n-1} B^2}{2(1\mp x)^{2n}}{\det (\bar{g}_{i\bar{j}})}
\ee
and therefore
\bea
\label{bexp}
&&b^{\pm}= \bar{b}+\frac{i}{2} \frac{\partial}{\partial w^{\pm} }\left[  \log \frac{A^{2(n-1)} B^2}{(1\mp x)^{2n}}  \right] \left[dw^{\pm} \pm (\partial -\bar{\partial})k \right]  \nonumber \\
&& \qquad - \frac{i}{2}\frac{\partial}{\partial \bar{w}^{\pm} }\left[  \log \frac{A^{2(n-1)} B^2}{(1\mp x)^{2n}}  \right] \left[ d\bar{w}^{\pm} \mp  (\partial -\bar{\partial})k \right]
\eea
where $\hat{\rho}^{\pm}=db^{\pm}$ and $\bar{\rho}=d\bar{b}$ are the Ricci forms of $\hat{\gamma}^{\pm}_{AB}$ and $\bar{g}_{ab}$ respectively. Note that to perform this calculation one must take care of the fact that by definition the complex coordinates $w^{\pm}$ are shifted by a function of $(z^i,\bar{z}^i)$ (i.e. $k$): while this does not change the $\partial/\partial w^{\pm}$ derivatives it does shift
\be
\frac{\partial}{\partial z^i} \to \frac{\partial}{\partial z^i} \pm  \; \partial_i k  \left( \frac{\partial}{\partial w^{\pm}} +\frac{\partial}{\partial \bar{w}^{\pm}} \right)  \; .
\ee
Noting that
\begin{equation}
\frac{\partial}{\partial w^{\pm}} = \frac{B^2}{2} \frac{\partial}{\partial x} \mp \frac{ i}{2} \frac{\partial}{\partial \phi}
\end{equation}
the final two terms of~(\ref{bexp}) can then be written back in our real coordinates:
\be
b^{\pm}= \bar{b} \mp \frac{ B^2}{2} \frac{d}{d x}\left[  \log \frac{A^{2(n-1)} B^2}{(1\mp x)^{2n}}  \right] (d\phi+\sigma)
\ee
Since the base is K\"ahler-Einstein $\bar{\rho}= 2n \bar{J}$ and therefore locally $\bar{b}= n\sigma$ up to an exact 1-form -- we will choose this to be such that $\bar{b}=n(d\phi+\sigma)$ and therefore
\be
b^{\pm}=\left( n \mp \frac{ B^2}{2} \frac{d}{d x}\left[  \log \frac{A^{2(n-1)} B^2}{(1\mp x)^{2n}}  \right]  \right) (d\phi+\sigma)  \; .
\ee
Using our explicit form for $B^2$ finally gives the Ricci form of $(\mathcal{H},\hat{\gamma}^{\pm})$
\be
\label{ricciform}
\hat{\rho}^{\pm}= d\left[f(x)(d\phi+\sigma)\right]
\ee
where
\be
\label{f}
f(x)=n \mp \frac{1}{2(1-x^2)^{n-1}(\xi+x^2)}\left( P'-\frac{2xP}{(\xi+x^2)} \pm \frac{2nP}{1\mp x} \right) \; .
\ee

\section{Topology of $\mathcal{H}_{m,p}$}\label{topH}
In the main text we established that our horizon metrics extend smoothly onto compact $\mathcal{H}_{m,p}$, and that the resulting $\mathcal{H}_{m,p}$ is the total space of an $S^2$-bundle over a compact K\"ahler-Einstein manifold $K$. Note that the structure group for such bundles is at most $SO(3)$. In this section we will discuss the topology of these bundles in more detail. To avoid clutter we will refer to $\mathcal{H}_{m,p}$ simply as $\mathcal{H}$.

 One of the most fundamental topological invariants of a manifold is the fundamental group. Since the fibre of the bundle $\mathcal{H}$ over $K$ is $S^2$, and $S^2$ is simply connected, it follows that $\pi_1(\mathcal{H})\cong \pi_1(K)$ \footnote{For any fibre bundle $F\to E \to B$, where $F$ is the fibre, $E$ the total space and $B$ the base, exactness of its homotopy sequence $\cdots \to \pi_1(F) \to \pi_1(E) \to \pi_1(B) \to \pi_0(F)\to \cdots$, implies that if $\pi_0(F)=\pi_1(F)=0$ then $\pi_1(E)$ is isomorphic $\pi_1(B)$, see e.g.~\cite{Stenrod, BottTu}.}. Furthermore since any closed K\"ahler manifold with positive definite Ricci tensor must be simply connected~\cite{Kob61} we must have $\pi_1(K)=0$ (recall $K$ is positive Einstein). Therefore
\be
\pi_1(\mathcal{H})=0 \; .
\ee
From this we immediately deduce that $H_1(\mathcal{H})=0$ and by Poincare duality the free part of $H_{2n-1}(\mathcal{H})=0$.

In fact we can easily deduce the whole cohomology ring $H^*(\mathcal{H})$ as follows. First note that there is a closed global 2-form $J_+/(1-x)^2$ on $\mathcal{H}$, see equation (\ref{Jpm}),  whose restriction to each fiber $S^2$ generates the cohomology of the fiber. Then, by the Leray-Hirsch theorem~\cite{BottTu}, the cohomology of $\mathcal{H}$ is
\begin{equation}
H^{\star}(\mathcal{H}) = H^{\star}(K) \otimes H^{\star}(S^{2}) \; .
\end{equation}
This allows us to deduce $H^{k}(\mathcal{H},\mathbb{R}) = H^{k}(K,\mathbb{R}) \oplus H^{k-2}(K,\mathbb{R})$ for $k \geq 2$.  Now, recall that the Euler characteristic $\chi(\mathcal{H})=\sum_{i=0}^{2n} (-1)^i b_i(\mathcal{H})$, where $b_i(\mathcal{H})=\textrm{dim}\, H^i(\mathcal{H},\mathbb{R})$ are the Betti numbers of $\mathcal{H}$. Since $b_{k}(\mathcal{H}) = b_{k}(K) + b_{k-2}(K)$ for $2\leq k \leq 2n-2$,  and $b_1=b_{2n-1}=0$ from above, we deduce
\begin{equation}
\chi(\mathcal{H}) = 2\chi(K)
\end{equation} for any choice of $K$.

For later use we now construct an explicit basis for the second integral homology groups $H_2(\mathcal{H},\mathbb{Z})$ in terms of that of $H_2(K,\mathbb{Z})$, following~\cite{GMSW}.   Since $\pi_1(K)=0$ we have $H^2(K,\mathbb{Z})\cong \mathbb{Z}^r$ for some $r$ (i.e. the torsion vanishes), and therefore we can find 2-cycles $\Sigma_i$ for $i=1, \cdots, r$ such that the homology classes $[\Sigma_i]$ form a basis for the free part of $H_2(K,\mathbb{Z})$.
Now define a submanifold $\Sigma \cong S^2$ of $\mathcal{H}$ corresponding to the fibre of $\mathcal{H}$ at some fixed point on $K$. Also define the global section $s: K \to \mathcal{H}$, of $\pi: \mathcal{H}\to K$,  by the property that each base point in $K$ is mapped to the pole $x=x_2$ of the fibre $S^2$. Therefore $\{ \Sigma, s\Sigma_i \}$ forms a representative basis for the free part of $H_2(\mathcal{H},\mathbb{Z})$. Observe that we have $H^2(\mathcal{H}, \mathbb{Z})= \mathbb{Z} \oplus H^2(K, \mathbb{Z})$ which agrees with the general result derived above.

Next we note that $\mathcal{H}$ is a complex manifold -- this is proved in the Appendix (\ref{appendix:kahler}). We can introduce complex coordinates $(W, z^i)$, where $W=\exp( 2\pi w_+/\Delta \phi)$ with
\be
w_+= i\phi -k(z,\bar{z}) +\int^x \frac{dx}{B^2}
\ee
and $z^i$ a set of complex coordinates inherited from $K$ where $k$ is the K\"ahler potential of $K$. Note that $W$ is well defined everywhere except at the pole $x=x_2$. Furthermore $W=0$ at the other pole $x=x_1=-x_2$. To see these facts one simply has to note that $B^2\sim (\const)^2 (x_2^2-x^2)$ as $x^2\to x_2^2$ and thus $\int^xdx/B^2 \to \infty$ as $x\to x_2$ and $\int^x dx/B^2 \to -\infty$ as $x \to -x_2$. Thus $W$ is a coordinate on the $\mathbb{CP}^1 \cong S^2$ fibre of $\mathcal{H}_{m,p}$ which covers everywhere except the $x=x_2$ pole. Hence $1/W$ is a coordinate on the fibre which covers everywhere except the other pole $x=x_1$.

Using the complex structure we can now relate our bundle $\mathcal{H}$ to a standard one. Let $\mathcal{L}$ denote the canonical line bundle over $K$ (i.e. the holomorphic line bundle of  $(n-1,0)$ forms over $K$). By adding a point to each fibre $\mathbb{C}$ one has an associated $\mathbb{CP}^1\cong S^2$ bundle over $K$, such that the $U(1)$ subgroup of transitions functions of $\mathcal{L}$ acts isometrically on $\mathbb{CP}^1$. For this canonical $S^2$ bundle $(\Delta \phi)_c = 2\pi /n$ (see e.g.~\cite{GMSW}). In our case we have $\Delta \phi= (p/m) (\Delta \phi)_c$. Therefore the coordinate on the $\mathbb{CP}^1$ fibre of $\mathcal{H}_{m,p}$ can be written as $W=W_c^{m/p}$ where $W_c=\exp(2\pi w_+/(\Delta \phi)_c)$ is a coordinate on the $\mathbb{CP}^1$ fibre of $\mathcal{L}$. It follows that our $S^2$ bundles may be thought of as being associated to the $(m/p)^{\textrm{th}}$ power of the canonical line bundle, i.e. $\mathcal{L}^{\otimes \frac{m}{p} }$. Once again the association involves simply adding a point to each fibre $\mathbb{C}$ and taking the same $U(1)$ transition function such that they act isometrically on the $S^2$ fibre. These bundles are always well defined since $\mathcal{L}^{1/p}$ is always well defined (e.g. for $\mathbb{CP}^{n-1}$ this is the tautological bundle). An easy way to see this is from equation (\ref{c1K}). The line bundle $\mathcal{L}^{1/p}$ may be defined by $<c_1(\mathcal{L}^{1/p}), [ \Sigma_i ] > = p^{-1} <c_1(\mathcal{L}), [ \Sigma_i ] >= - p^{-1} < c_1(K), [\Sigma_i ] > = -n_i \in \mathbb{Z}$. As in~\cite{GMSW}, we may therefore write $\mathcal{H}_{m,p} \cong \mathcal{L}^{m/p} \otimes_{U(1)} \mathbb{CP}^1$,
which means take the same $U(1)$ transition functions as for $\mathcal{L}^{m/p}$ and use them to construct an associated $\mathbb{CP}^1\cong S^2$ bundle with the $U(1)$ acting isometrically on the fibre.  It is also worth noting that our bundle can be written as a projectivised bundle.

We are interested in determining the conditions for $\mathcal{H}$ to be a spin manifold. Recall this requires one to compute the second Stiefel-Whitney class $w_2(\mathcal{H})$. Since $T\mathcal{H}$ is a complex vector bundle we can use the fact that $w_2(\mathcal{H})$ is the mod $2$ reduction of the first Chern class $c_1(\mathcal{H})$.
Now we remark that our $(\mathcal{H},\gamma)$ is a conformally K\"ahler manifold  -- this property is shown in the Appendix \ref{appendix:kahler}. For definiteness we will chose the K\"ahler metric to be $\hat{\gamma}_{AB}=(1-x)^{-2}\gamma_{AB}$ (see Appendix \ref{appendix:kahler}).  Then, since the Chern class $c_1(\mathcal{H})$ (i.e. the first Chern class of the complex tangent bundle $T\mathcal{H}$) is a topological invariant, it is in fact easier to calculate this using the K\"ahler metric $\hat{\gamma}_{AB}$. This is a standard result for K\"ahler manifolds and is given by $c_1(\mathcal{H})= [ \hat{\rho} /2\pi ]$ where $\hat{\rho}$ is the Ricci form of $\hat{\gamma}_{AB}$.  The Ricci form $\hat{\rho}$ is given by (\ref{ricciform}). We can now evaluate the first Chern class evaluated on our basis of  the free part of $H_2(\mathcal{H},\mathbb{Z})$:
 \bea
 &&< c_1(\mathcal{H}), [\Sigma] >= \frac{1}{2\pi} \int_{\Sigma} \hat{\rho} = \frac{\Delta \phi}{2\pi} (f(x_2)-f(x_1)) =  2\\
 &&<c_1(\mathcal{H}), s_*[\Sigma_i]>= \frac{1}{2\pi}\int_{s\Sigma_i} \hat{\rho} = \frac{f(x_2)}{2\pi} \int_{\Sigma_i} 2J=  n_i(m+p)
\eea
where the second equality follows from the form of $f(x)$ (\ref{f}), $x_1 = -x_2$, the identity
\be
f(\pm x_2) = n \pm \frac{ 2\pi}{\Delta \phi}
\ee
(which can be derived using (\ref{Pder}) and (\ref{deltaphi})),  equation (\ref{2Ji})  and (\ref{deltaphim}).  Note that $n_i \in \mathbb{Z}$ are the same integers as in (\ref{c1K}) and satisfy gcd$(n_i)=1$.

We are now in a position to deduce the second Stiefel-Whitney class using $w_2(\mathcal{H}) =c_1(\mathcal{H}) \; \textrm{mod} \; 2$. This means that evaluating $w_2(\mathcal{H})$ on $H_2(\mathcal{H},\mathbb{Z})$ gives a set of integers which are the same mod $2$ as evaluating $c_1(\mathcal{H})$ on $H_2(\mathcal{H},\mathbb{Z})$. Therefore we deduce that $w_2(\mathcal{H})$ is trivial, and hence $\mathcal{H}$ is a spin manifold, if and only if $m + p$ is even. Note that this does not depend on whether $w_2(K)$ is trivial which occurs if and only if $p$ is even (see (\ref{c1K})).

Finally, we show that for $n\geq 3$ the bundle $S^2\to  \mathcal{H} \to K$ is never trivial. This is in contrast to the case $n=2$, which as we saw is the trivial bundle $S^2\times S^2$ if $m$ is even. To prove this we will use the fact that $S^2$-bundles over a compact manifold are partially classified by the first Pontryagin class of the associated $\mathbb{R}^3$-bundle (constructed with the same $SO(3)$ transition functions).  Explicitly, $\mathcal{H}_{m,p}$ may be thought of as the unit sphere bundle in $V_3= I_{\mathbb{R}} \oplus \mathcal{L}^{-m/p}$ where $I_{\mathbb{R}}$ is the trivial real line bundle over $K$. Therefore we need $p_1(V_3)= -c_2(V_3 \otimes \mathbb{C}) \in H^4(K, \mathbb{Z})$,  where $V_3\otimes \mathbb{C}= I_{\mathbb{C}}\oplus \mathcal{L}^{-m/p} \oplus \mathcal{L}^{m/p}$ and $I_{\mathbb{C}}$ is the trivial complex line  bundle over $K$. The total Chern class $c(V_3\otimes \mathbb{C})=c(\mathcal{L}^{-m/p})c(\mathcal{L}^{m/p})=(1-c_1(\mathcal{L}^{m/p}))(1+c_1(\mathcal{L}^{m/p}))=1-c_1(\mathcal{L}^{m/p})^2$. It follows that $p_1(V_3)=m^2 c_1(\mathcal{L})^2/p^2=m^2c_1(K)^2/p^2$. Since $K$ is K\"ahler-Einstein $c_1(K)= n[J]/\pi$ and thus finally we have
\be
p_1(V_3)= \frac{m^2 n^2}{\pi^2 p^2} [J]^2 \; .
\ee
This immediately implies that $p_1(V_3)\neq 0$ (since for our solutions $m \neq 0$) and therefore all our bundles are non-trivial -- of course this argument only works for $n\geq 3$. In fact we may go further and define a topological invariant for any $S^2$-bundle over a compact manifold, by the scalar quantity\footnote{We thank James Sparks for pointing this out.}
\be
\rho(\mathcal{H}) \equiv \int_K p_1(V_3) \wedge J^{n-3} = \frac{m^2 n n!}{p^2 \pi^2} \textrm{vol}(K)
\ee
which of course is only defined for $n \geq 3$. This number is an invariant of our bundles $\mathcal{H}_{m,p}$ -- in particular any two $S^2$-bundles over $K$ are not homeomorphic if the invariant $\rho$ just defined is different. We deduce that for different $m>0$ the manifolds $\mathcal{H}_{m,p}$ are not homeomorphic. This is in marked contrast to the $n=2$ case. Thus since we can have $m>p$  we in fact have an infinite discrete family of topologies (this is in contrast to the compact Einstein spaces in~\cite{PP} which have $m<p$).

\section{Inhomogeneous $\mathbb{CP}^n$ horizon with conical singularity} \label{caseII}
Here we perform the global analysis of case II in table 1, and without loss of generality take $-1<x_1\leq x \leq 1$ with $P(x_1)=0$ and $P(1)=0$ and $P(x)>0$ for $x_1<x<1$, so we must have $\epsilon=1$. This form of $P(x)$ clearly requires $P'(x_1)>0$ and comparing to equation (\ref{Pder}) implies $x_1<0$.

One can solve the constraint $P(1)=0$ to get
\be
c=c_*=-\frac{n! \sqrt{\pi}(2n-1-\lambda)}{\Gamma(n+\frac{1}{2})} (\xi-\xi_*)
\ee
where $\xi_*$ is given by (\ref{xiMP}). Substituting back gives\footnote{Note that this form for $P(x)$ is guaranteed by (\ref{Peq}) which allows one to show that if $P(1)=0$ then $P^{(m)}(1)=0$ for $1 \leq m \leq n-1$.}
\be\label{Pcase2}
P(x)=(1-x)^n R(x)
\ee
where $R(x)$ is a polynomial of order $n+2$ which we do not need explicitly for our analysis below.
However,  we do require the polynomial $R(x)$ to have a root $-1<x_1<0$ such that $R(x)>0$ for $x_1 < x < 1$.
The horizon metric in this case is
\be
\gamma_{AB}dx^A dx^B = L^2\left[  \frac{(\xi+x^2)(1+x)^{n-1}dx^2}{(1-x)R(x)} + \frac{(1-x)R(x)}{(\xi+x^2)(1+x)^{n-1}} (d\phi+\sigma)^2+ (1-x^2) \bar{g} \right]  \; .
\ee
Let us now examine regularity of this metric which is clearly non-degenerate and smooth for $x_1<x<1$. Near $x=1$ set $x=1-\eta^2$ and expand for small $\eta$. To leading order one gets
\be
\gamma_{AB}dx^Adx^B \sim  2L^2 \left[ d\eta^2 + \eta^2( (d\phi+\sigma)^2+\bar{g}) \right]
\ee
where we have used the identity\footnote{This may be obtained by differentiating~(\ref{Pcase2}) $n$ times, evaluating the result at $x=1$, and comparing to the expression obtained from computing $d^{n}P(x)/dx^n$ directly from~(\ref{Peq}). } $R(1)=2^n(1+\xi)$. Therefore smoothness at $\eta=0$ requires $\Delta \phi=2\pi$ and $\bar{g}$ to be the Fubini-Study metric on $\mathbb{CP}^{n-1}$. At the other endpoint $x_1$ we have $A^2>0$ and thus we have a bolt. Smoothness requires the conical singularity at $x=x_1$ to be removed.  The condition for this is (\ref{period}) (recall $x_1<0$ in this case)
\be
\Delta \phi= \frac{4\pi |x_1|}{2n-\lambda(1-x_1^2)}
\ee
and therefore since we have already shown $\Delta \phi=2\pi$, smoothness requires
\be
\label{CPnsing}
|x_1|=n-\frac{\lambda}{2}(1-x_1^2)\; .
\ee
Recall that the root $x_1$ must satisfy $-1<x_1<0$. Therefore for $\lambda \leq 0$ the regularity condition (\ref{CPnsing}) implies $x_1\leq -n$ which is a contradiction. Therefore for $\lambda \leq 0$ the above metric necessarily is singular (either at $x=1$ of $x=x_1$). It can be thought of as a metric on $\mathbb{CP}^n$ with a conical singularity at the bolt.

It is possible that for $\lambda>0$ this metric can be made smooth -- to prove this one needs to show that $R(x)$ has a root $x_1$ such that $-1<x_1<1$, $R(x)>0$ for $x_1 < x \leq 1$ and (\ref{CPnsing}) is satisfied. Then the horizon metric would be a smooth and inhomogeneous metric on $\mathbb{CP}^n$. We will not pursue this here.

\section{Computation of  ``internal" angular momenta on $K$}\label{IntJ}
In this section we compute the Komar integral (\ref{Jkomar}) associated to the $U(1)$ Killing vector fields $\bar{m}_i = \partial/\partial \phi^i$ for $i=1, \cdots, n-1$ for $K$ toric. Explicitly this is
\begin{equation}
\bar{J}_i = \pm \frac{L^{2n}\Delta\phi \sqrt{\xi}}{8\pi G} \int_{-x_2}^{x_2} \frac{P(x)}{(\xi + x^2)^2} dx \int_{K} (\sigma \cdot \bar{m}_i )  \; \bar{\epsilon}
\end{equation} where $\bar{\epsilon}$ is the volume form on $K$.  The calculation thus reduces to evaluation of the integrals
\be
I_i \equiv \int_{K} (\sigma \cdot \bar{m}_i )  \; \bar{\epsilon}   \; .
\ee
First observe that this integral is actually well defined despite $\sigma$ not being a globally defined object on $K$. To see this recall the K\"ahler form $J=\frac{1}{2} d\sigma$ and $\mathcal{L}_{\bar{m}_i} J=0$. It is therefore possible to choose a gauge for $\sigma$ such that $\mathcal{L}_{\bar{m}_i} \sigma=0$. Any residual gauge transformations $\sigma \to \sigma + d\lambda$ must satisfy $\mathcal{L}_{\bar{m}_i} d\lambda=0$ which is equivalent to the function $\bar{m}_i \cdot d\lambda = \const$. However this constant must vanish since $\bar{m}_i$ each vanish somewhere on $K$. This proves that $I_i$ is gauge invariant and thus well defined. We will now do an explicit calculation for $K=\mathbb{CP}^{n-1}$ and also give a more general argument for toric $K$.

\subsection{$K = \mathbb{CP}^{n-1}$}
The calculation thus reduces to evaluation of the integrals
\be
I_i \equiv \int_{\mathbb{CP}^{n-1}} (\sigma \cdot \bar{m}_i )  \; \bar{\epsilon}   \; .
\ee
As noted above $I_i$ is gauge invariant. Therefore it may be computed by choosing an open covering of $\mathbb{CP}^{n-1}$ and working in a gauge there $\sigma$ is smooth in each open set.

Now recall the standard open cover of $\mathbb{CP}^{n-1}$ which consists of $n$ patches $U_k=\{ Z_k \neq 0 \}$ with $k=1, \cdots, n$ and $Z_k$ the usual homogeneous coordinates. In each patch $U_k$ we can introduce inhomogeneous coordinates $z^{(k)}_i=Z_i/Z_k$ for $i \neq k$. Let us work in one patch say $U_n$ and set $z^{(n)}_i=z^i$ for $i=1, \cdots, n-1$. The Fubini-Study metric in such a patch is given by
\begin{equation}\label{FSmet}
\bar{g}_{ab} d\bar{x}^a d\bar{x}^b= d\Sigma_{n-1}^2 = \frac{dz^i d\bar{z}^{{i}}}{f}- \frac{\bar{z}^{{i}} z^{j} dz^i d\bar{z}^{{j}}}{f^2}
\end{equation} where $f = 1 + z^i \bar{z}^{{i}}$ and $i,j=1, \cdots, n-1$ and we are summing over repeated indices (this metric satisfies ${\textrm{Ric}(\bar{g})} = 2n \bar{g}$). The K\"ahler form is
\begin{equation}
J = \frac{d\sigma}{2}, \qquad \sigma = \frac{i}{2f}\left(z^i d\bar{z}^{{i}} - \bar{z}^{{i}}dz^i\right) \; .
\end{equation} There are a number of ways to introduce real coordinates in order to identify the rotational Killing fields $\bar{m}_i$. The simplest for our purposes is to set $z^i = r_i e^{i\phi^i}$. The vector fields $\bar{m}_i= \partial /\partial  \phi^i$ have closed orbits with period $2\pi$ and generate the $U(1)^{n-1}$ isometry subgroup, and $0 \leq r_i \leq \infty$.  Then (\ref{FSmet}) becomes
\begin{equation}
d\Sigma^2_{n-1} = \frac{1}{f} \left[ \sum_{i=1}^{n-1}  dr_i^2 + r_i^2(d\phi^i)^2 \right]  - \frac{1}{f^2}  \left[ \left( \sum_{i=1}^{n-1}r_i dr_i   \right)^2 + \left(\sum_{i=1}^{n-1}r_i^2 d\phi^i \right)^2 \right]
\end{equation} where $f= 1 + \sum_{i=1}^{n-1} r_i^2$. It follows that
\begin{equation}
\sqrt{\bar{g}} = \frac{\prod_{i=1}^{n-1} r_i}{f^n}  \; .
\end{equation}
We also have
\begin{equation}
\sigma = \frac{1}{f}\sum_{i=1}^{n-1} r_i^2 d\phi_i  \; .
\end{equation}
Note this is in a gauge which ensures $\sigma$ is smooth at $r_i=0$. Putting this together gives
\begin{eqnarray}
I_i^{(n)} \equiv  \int_{U_n}(\sigma \cdot \bar{m}_i )  \; \bar{\epsilon}  &=& (2\pi)^{n-1}\int_0^\infty dr_1 \int_0^\infty dr_2 \ldots\int_0^{\infty}dr_{n-1} \frac{r_i^3 \prod_{j=1, j\neq i}^{n-1} r_j}{f^{n+1}} \nonumber \\
 &=& \frac{(2\pi)^{n-1}}{2^{n-3} n!}\int_0^{\infty} \frac{r_i^3 \; dr_i}{(1+r_i^2)^3} = \frac{ \pi^{n-1}}{n!}
 \end{eqnarray} by repeated integration.

 Now, it is clear that the analogous integrals  $I^{(k)}_i$ in the patches $U_k$ for $k=1, \cdots, n-1$ give the same value. However, it is not the case that adding all these together gives $I_i$, since the overlaps $U_i \cap U_j \neq 0$. In fact there is  a trick to avoid this complication. Instead one can work in a gauge which is singular in every patch, in such a way that performing the integral in any patch gives the correct {\it total} answer. For the case at hand this gauge is given by
 \be
 \sigma = \sum_{i=1}^{n-1}\left(  \frac{r_i^2}{f} -\frac{1}{n}  \right) d\phi_i
\ee
which gives
\be
I_i=I^{(n)}_i -\frac{1}{n} \int_{\mathbb{CP}^{n-1}} \bar{\epsilon}  =0
\ee
where in the last equality we have used $\textrm{vol}(\mathbb{CP}^{n-1})= \frac{\pi^{n-1}}{(n-1)!}$.
Hence as expected, the conserved angular momenta associated with the `internal' rotational Killing fields on $\mathbb{CP}^{n-1}$ vanish.

\subsection{Toric $K$}
In this section we generalise the calculation of the previous section to cover the general case of when $K$ is a toric manifold. As we will see, we do not actually need the explicit metric in order to calculate the internal angular momenta, just the toric data. We will employ well known constructions of toric symplectic geometry.

First recall that for a $2(n-1)$ dimensional toric K\"ahler manifold one may introduce symplectic coordinates $(x^i, \phi^i)$ for $i=1, \cdots n-1$ such that the metric is
\be
\bar{g}_{ab}d\bar{x}^a d\bar{x}^b =  G_{ij}(x) dx^i dx^j + G^{ij}(x) d\phi^i d\phi^j
\ee
where $G^{ij}$ is the matrix inverse of $G_{ij}$, and  $G_{ij}= \partial^2 g/ \partial x^i \partial x^j$ where $g$ is called the symplectic potential. The symplectic form is the K\"ahler form which is
\be
J= dx^i \wedge d\phi^i
\ee
so in the language of symplectic geometry these are Darboux coordinates.
The coordinates ranges are $\phi^i \sim \phi^i +2\pi$, so that the Killing vector fields $\bar{m}_i =\partial / \partial \phi^i$ generate the toric symmetry $U(1)^{n-1}$. By a classic result the $x^i$ are coordinates which lie in a so called Delzant polytope $\Delta$. This is a subset of $\mathbb{R}^{n-1}$ defined by the intersection of a set of linear inequalities $\Delta= \{ x:  (v_a \cdot x +\lambda_a) \geq 0 \; ,\;\forall a \}$, where $a$ labels the faces of the polytope and $v_a$ is the normal vector to each face such that $v_a$ form a basis for $\mathbb{Z}^{n-1}$. Note that symplectic coordinates are not unique. In particular $x^i \to M^i_{\phantom{j}j} x^j$ where $M\in GL(n-1, \mathbb{Z})$ and $ x^i\to x^i+k^i$ are both freedoms. The polytope $\Delta$ is then invariant under a subgroup of these transformations.

As in the previous section the calculation of the internal angular momenta reduces to
\be
I_i= \int_K (\sigma \cdot \bar{m}_i ) \; \bar{\epsilon}  \; .
\ee
In symplectic coordinates we can always choose a gauge such that
\be
\sigma= 2 (x^i+c^i) d\phi^i
\ee
for some constants $c^i$.
Also note that the volume form in these coordinates is trivial so $\sqrt{\bar{g}}=1$. Therefore
\be
I_i= 2 (2\pi)^{n-1} \int_{\Delta} (x^i+c^i) \; dx^1 \cdots dx^{n-1}  \; .
\ee
It is then clear that we can always pick a gauge such that $I_i=0$ for all $1 \leq i \leq n-1$, i.e. $c^i=-\textrm{vol}(\Delta)^{-1} \int_{\Delta} x^i \; dx^1 \cdots dx^{n-1}$. By the same reasoning as in the previous section, since $I_i$ is gauge invariant, if it vanishes in every coordinate patch then it must vanish everywhere. Thus we deduce that for general toric $K$ the internal angular momenta vanish.

\end{document}